\documentclass[11pt,english]{article}
\usepackage[utf8]{inputenc}
\usepackage{amsmath}
\usepackage{amsthm}
\usepackage{amssymb}
\usepackage{graphicx}

\makeatletter

\DeclareTextSymbolDefault{\textquotedbl}{T1}

\@ifundefined{date}{}{\date{}}
\usepackage{t1enc}
\usepackage[english]{babel}
\usepackage{amsthm}
\usepackage{braket}
\usepackage{centernot}
\usepackage{tikz}

\def\cros{\raise1.9pt\hbox{$\scriptscriptstyle
          >$}\!\raise1.5pt\hbox{$\scriptstyle\triangleleft\,$}}

\def\l{{\lambda}}

\theoremstyle{definition}\theoremstyle{definition}\theoremstyle{definition}\theoremstyle{definition}

\title{\bf Operational equivalence \\ and causal structure}
\author{\textit{Gábor Hofer-Szabó}\thanks{HUN-REN Research Center for the Humanities, Budapest, email: szabo.gabor@btk.mta.hu}}

\makeatother

\usepackage{babel}

\begin{document}
\maketitle
\abstract{In operational quantum mechanics two measurements are called operationally
equivalent if they yield the same distribution of outcomes in every
quantum state and hence are represented by the same operator. In this
paper, I will show that the ontological models for quantum mechanics
and, more generally, for any operational theory sensitively depend
on which measurement we choose from the class of operationally equivalent
measurements, or more precisely, which of the chosen measurements
can be performed simultaneously. To this goal, I will take first three
examples---a classical theory, the EPR-Bell scenario and the Popescu-Rochlich
box; then realize each example by two operationally equivalent but
different operational theories---one with a trivial and another with
a non-trivial compatibility structure; and finally show that the ontological
models for the different theories will be different with respect to
their causal structure, contextuality, and fine-tuning. }

\section{Introduction: a Bridgmanian perspective}

On strict operationalism, concepts should be defined by empirical
operations. In this tradition, going back to Percy Bridgman (1927)
and the Vienna Circle (Schlick, 1930), two concepts which are defined
by different operational procedures cannot be the same. Using Bridgman's
example, length measured by a ruler and length measured by light signals
are different concepts, and true science should use different names
to discern them. As times passed, philosophy of science (and also
Bridgman himself) has gradually moved away from strict operationalism
and revealed various semantic, pragmatic and common sense criteria
for identifying concepts with different operational basis (Chang,
2019). In the case of physical magnitudes or observables, the standard
way was to check whether the two measurements defining the two observables
have the same outcome in their common domain. If the length of medium
sized objects agree when measured by a ruler or measured by light
signals, then---at least in this common domain---one is justified
in using one length concept instead of two.

For this comparison, however, at least one of the two conditions needs
to hold for each system: 
\begin{center}
\begin{enumerate}
\item[(i)] either both measurements should be able to be performed \textit{simultaneously} on the system; or 
\item[(ii)] we need to have a precise enough preparation procedure such that every preparation is an \textit{eigenstate} for both measurements, that is a preparation in which  both measurements have definite result.
\end{enumerate}
\par\end{center}

This latter happens in classical physics where each pure (dispersion-free)
state is an eigenstate for any measurements. Thus in classical physics,
we can easily decide on whether two measurements measure the same
observable: just prepare the system in an eigenstate, perform the
one measurement, prepare the system again in the same eigenstate,
perform the other measurement and compare the outcomes. However, if
the preparation procedure is not as fine-grained as to yield definite results
for all measurements, as is the case in quantum mechanics, we are
left with option (i) to identify observables of different measurements:
we need to measure them simultaneously and check whether the outcomes
match in every single run. 

But what if the two measurement procedures cannot be performed at
the same time? From a strict operationalist position, we are not entitled
to identify the two observables in this case. Still, in quantum mechanics
this is what happens. Here, two measurements which yield the same
outcome statistics in every quantum state are represented by the same
operator and said to measure the same observable. Such measurements
are called \emph{operationally equivalent}. From the Bridgmanian perspective,
the identification of the observables of operationally equivalent
measurements is physically unjustified. The mere statistical match
of outcomes of two measurements which cannot be performed at the same
time on the same system does not guarantee that the two measurements
would give the same outcome run-by-run and hence that they measure
the same observable. 

Operational equivalence is sometimes expressed in the form that observables
are not associated with measurement procedures, as in Bridgman, but
with operators. So instead of \emph{one measurement--one observable}
we have \emph{one operator--one observable}. Let me refer to the
first identification of observables as \emph{Bridgmanian} and to the
second as \emph{standard} (standard in quantum mechanics). Schematically:
\begin{center}
\begin{tabular}{rccccccc} &&&&&&&\\ 
  &  & \textit{Bridgmanian} &  &  &  & \textit{Standard} &  \\
  &   &  &  &  &  &  &  \\ 
\textit{Operator:} \quad &  & $\textbf{O}$ &  &  &  & $\textbf{O}$ &  \\
  &   &  $\swarrow \quad \searrow$ &   &  &  & $\downarrow$ &  \\  
\textit{Observable:} \quad & & $\mathcal{O}_1 \quad \quad \quad \mathcal{O}_2$   &  &  &  & $\mathcal{O}$ &  \\
  &  & $\downarrow \quad \quad \quad \quad \downarrow$   &  &  &  & $\swarrow \quad \searrow$ & \\  
\textit{Measurement:} \quad &  & $M_1 \quad \quad \quad M_2$ &  &  & & $M_1 \quad \quad \quad M_2$ & \\
  &  &  & &  &  &  & 
\end{tabular}
\par\end{center}

As an example, consider the following two measurement procedures for
photon polarization. ``The first, which we denote by $M_{1}$, constitutes
a piece of polaroid oriented to pass light that is vertically polarized
along the $\hat{z}$ axis, followed by a photodetector. The second,
which we denote by $M_{2}$, constitutes a birefringent crystal oriented
to separate light that is vertically polarized along the $\hat{z}$
axis from light that is horizontally polarized along this axis, followed
by a photodetector in the vertically polarized output.'' (Spekkens,
2005, p. 2). The two measurements are operationally equivalent: they
provide the same distribution of outcomes for photons in any quantum
state. Consequently, they are represented by the same operator, $\boldsymbol{\sigma}_{z}$,
in quantum mechanics. 

Now, do $M_{1}$ and $M_{2}$ measure the same observable or they
measure different observables?\footnote{Note that Spekkens, as a good operationalist, does not use the term
''observable'' in his 2005 paper.} According to the standard quantum mechanical approach, they measure
the same observable $\mathcal{O}$, the polarization of the photon.
According to the Bridgmanian approach, they measure different observables:
$M_{1}$ measures $\mathcal{O}_{1}$, the polarization of the photon
with respect to a piece of polaroid, and $M_{2}$ measures $\mathcal{O}_{2}$,
the polarization of the photon with respect to a birefringent crystal. 

Note that the above Bridgmanian conditions to identify $\mathcal{O}_{1}$
and $\mathcal{O}_{2}$ are not fulfilled now: (i) $M_{1}$ and $M_{2}$
cannot be simultaneously measured on the very same photon since we
need to decide on whether we insert a polaroid or a birefringent crystal
in the path of the photon. (ii) The outcome of $M_{1}$ and $M_{2}$
will not necessarily will be the same for two photons prepared in
the same state if this state is not an eigenstate of $\boldsymbol{\sigma}_{z}$.
Thus, from the Bridgmanian perspective, the two observable should
not be identified. 

Even though from the Bridgmanian perspective the standard position
is unsatisfactory, it has its own rationale. If all that quantum mechanics
can predict is the distribution of outcomes, and if there are no preparations
which would discern two measurements with respect to the outcome distribution,
then why would one like to discern the two measurements? They measure
the same observables, like gas thermometer and alcohol thermometer
measure the same temperature, and any difference in the concrete realization
of the measurements is just of secondary importance.\footnote{Interestingly, we can defend the standard position even from a Bridgmanian
perspective \emph{when modifying the concept of measurement}. If we
take a measurement in quantum mechanics not to be a single measurement
but a sequence of measurements and an outcome not to be a single outcome
but a statistical distribution of outcomes, then we can apply criterion
(ii) in arguing that two sequences of operationally equivalent measurements
measure the same observable: they have the same distribution of outcomes
in every state, therefore they measure the same observable.}

In any event, the standard position remains consistent as long as
we remain at the level of quantum theory. But at the moment when we
try to extend the ontology by ontic (hidden) states, the identification
of observables corresponding different measurements represented by
the same operator becomes problematic. The Kochen-Specker theorems
highlight just this fact. It is instructive to see how Kochen-Specker
theorems are interpreted on the standard approach (Held 2022, Spekkens
2005, Hofer-Szab\'o, 2021a, b, 2022). On this account, the lesson of
the Kochen-Specker theorems is that the value of certain observables
associated with operators depends on the measurement with which it
is measured or co-measured. This fact is commonly referred to as contextuality---or
``ontological contextuality'' (Redhead, 1989) if not only the value
but also the observables themselves depend on which measurement they
are measured by. But note that from the Bridgmanian perspective there
is nothing contextual in this fact; it simply shows that we were too
quick to identify observables measured by different measurements when
we relied simply on the match of the outcome statistics.

\vspace{.1in} In this paper, I will revisit the Bridgmanian view of operationalism
and investigate how far we get when we do \emph{not} identify observables
associated with operationally equivalent measurements. To this goal,
I will use the framework of operational theories and ontological models
introduced by Rob Spekkens (2005). This framework is general enough
to embrace classical, quantum, super-quantum theories, and to analyze
contextuality, causal structure and many other important features
across the different theories. The main claim of the paper can be
formulated at the more specific level of quantum mechanics and at
the more general level of operational theories. As for quantum mechanics,
this claim reads as follows:

\vspace{.1in} \emph{Ontological models for quantum mechanics are sensitive
not only to the operators but also to the measurements realizing these
operators; more specifically, to whether these measurements can be
performed simultaneously or not. The ontological models for these
different measurements realizing the same set of operators in quantum
mechanics but having a different compatibility structure can be highly
different with respect to the causal structure, contextuality, and
fine-tuning.}

\vspace{.1in} This strong dependence of the properties of the ontological
models on the realizing measurements, however, is not restricted to
quantum mechanics. It is a general feature of any operational theory.
To show this, in the paper I will construct for an operational theory
another operational theory with a different compatibility structure
such that the measurements of the two theories are operationally equivalent,
still they admit different ontological models. Thus, the main general
claim of our paper is the following:

\vspace{.1in} \emph{Ontological models for general operational theories are
sensitive not only to the operationally equivalent classes of measurements
but also to the measurements themselves, more specifically, to the
compatibility structure of these measurements. Ontological models
for operationally equivalent theories with different compatibility
structures can be highly different with respect to the causal structure,
contextuality, and fine-tuning.}

\vspace{.1in} More specifically, I will do the following. Operational theories
come together with a set of measurements and a set of simultaneous
measurements. For any operational theory, I will construct another
theory with the following properties: a) the measurements of the new
theory are operationally equivalent to the measurements of the old
theory; b) in the new theory, there are no simultaneous measurements.
I will call this procedure\emph{ trivialization.} With this procedure
in hand, I will show the following:
\begin{enumerate}
\item The trivialization of an operational theory can be nicely represented
graph theoretically as taking the line graph of the graph representing
the original theory. 
\item On the example of three non-disturbing (no-signaling) operational
theories---a classical theory, the EPR-Bell scenario, and the Popescu-Rorhlich
box, I will show how the most important features of the ontological
models change when we replace an operational theory with a new, trivialized
theory. 
\item I will discern two different and logically independent concepts of
contextuality, simultaneous contextuality and measurement contextuality,
and show that the trivialization can alter the ontological models
with respect to the former but not to the latter.
\end{enumerate}
In the paper I will proceed as follows. After introducing the framework
of operational theories (Sec. \ref{Sec:optheor}) and ontological
models (Sec. \ref{Sec:ontmod}), I define the procedure of trivialization
(Sec. \ref{Sec:trivialization}). Next, I compare the ontological
model of three non-trivial (Sec. \ref{Sec:nontrivi}) and three corresponding
trivial (Sec. \ref{Sec:trivi}) operational theories. I analyze the
causal structure of the models (Sec. \ref{Sec:causal}), show how
trivialization leads to trivial causal graphs (Sec. \ref{Sec:Reptrivi}),
revisit the special case of quantum mechanics (Sec. \ref{Sec:QMech})
and conclude with a short discussion (Sec. \ref{Sec:Conclu}).

\section{Operational theories}

\label{Sec:optheor}

The concept of measurements can be analyzed from several directions (Tal, 2020). Mathematical theories of measurement (Suppes, 1951) are concerned with the question of how to assign abstract terms to physical magnitudes. Operationalism and conventionalism, on the other hand, focus on the semantics of measurements, and defines the meaning of quantity-concepts in terms of operations (Bridgman, 1927) or 'coordinative definitions' (Reichenbach 1927). Realism addresses the metaphysical nature of measurable quantities and conceives of measurement results as approximations of the true values of physical quantities (Trout 1998). Information-theoretic and model-based accounts examine the epistemological aspects (informativity, coherence, consistency) of measurements (van Fraassen, 2008).

In this paper, I will take an operationalist perspective to measurements. This perspective has a long tradition in quantum theory starting with von Neumann (1932) and followed by Mackey (1957), Ludwig (1983), Busch et al. (2016), D'Ariano et al. (2017) and many others. Note, however, that, unlike many in the mathematical physics community, I will use the term 'measurement' in the original operationalist meaning referring to the physical procedure itself and not to its various mathematical representations, such as self-adjoint operators, PVMs, POVMs, effects, or whatever. My approach will follow the operational theories and ontological models framework of Rob Spekkens (2005). In this and the next section, I introduce the main concepts of this framework. 

An \textit{operational theory} is a theory which specifies the probability
of the outcomes of certain measurements performed on a physical system
which was previously prepared in certain states. Let $\mathcal{P}=\{P_{1},P_{2},\dots\}$
be set of \textit{preparations} of the system, $\mathcal{M=}\{M_{1},M_{2},\dots\}$
the set of \textit{measurements} which can be performed on the system,
and let $\mathcal{X}=\{X_{1},X_{2},\dots\}$ be the set of \textit{outcomes}.\footnote{Without loss of generality, we can assume that all (basic, see below)
measurements have the same set of outcomes. If not, we just add null-outcomes
to the outcome set of some measurements.  } Let $P,M,$ and $X$ be random variables running over the preparations,
measurements and outcomes, respectively, assigning to each event its
index. (Thus, ``$P=1$'' refers to the preparation $P_{1},$ ``$M=2$''
refers to the measurement $M_{2},$ etc.) Using these random variables,
an operational theory is simply a set of \textit{conditional probabilities}
of the outcomes given the various measurements and preparations, that
is
\begin{eqnarray}
p(X|M,P)\label{optheory}
\end{eqnarray}
where $P,M,$ and $X$ run over the set $\mathcal{P}$, $\mathcal{M}$,
and $\mathcal{X}$, respectively.

Two measurements $M_{1}$ and $M_{2}$ are \textit{simultaneously
measurable}, if they can be performed on the same system at the same
time. Simultaneous measurability is an empirical question. Operationally,
one identifies measurements by sets of laboratory instructions. The
spin measurement of an electron, for example, is given by the detailed
description of the path of the electron, the position of the Stern-Gerlach
magnets and detectors, etc. As a consequence of this characterization
of measurements by sets of laboratory instructions, two measurements
$M_{1}$ and $M_{2}$ will be simultaneously measurable if and only
if there is a measurement which can be identified by the \textit{conjunction}
of the sets of instructions characterizing $M_{1}$ and $M_{2}$.
We call this measurement the \textit{simultaneous measurement} of
$M_{1}$ and $M_{2}$ and denote it by $M_{1}\wedge M_{2}$ (which is
again a measurement in $\mathcal{M}$). The random variable $M$ will
assign to $M_{1}\wedge M_{2}$ the pair $(1,2)$ and the outcomes of $M_{1}\wedge M_{2}$
are taken from the set $\mathcal{X}^{(1)}\times\mathcal{X}^{(2)}$.
From the definition of simultaneous measurements it also follows that
$M_{1}\wedge M_{2}\wedge M_{3}\in\mathcal{M}$ implies $M_{1}\wedge M_{2}\in\mathcal{M}$.
If $M_{1}$ and $M_{2}$ are not simultaneously measurable, we write
$M_{1}\wedge M_{2}\notin\mathcal{M}$. If a measurement in an operational
theory is not a simultaneous measurement of two or more other measurements,
then we call it a \emph{basic measurement}. 

Note that $M_{1}\wedge M_{2}$ and $M_{1}$ are simultaneous measurements
since the \textit{\emph{conjunction}} of the sets of instructions
characterizing $M_{1}\wedge M_{2}$ and $M_{1}$ is just the set characterizing
$M_{1}\wedge M_{2}$. Similarly, a measurement and a certain \emph{marginalization}
(see below) of this measurement are simultaneous measurements since
this latter measurement is just the measurement plus some extra instructions. 

An important consequence of defining measurements by sets of instructions
is that we do \emph{not} identify two measurements just because they
are operationally equivalent. Two measurements $M_{1}$ and $M_{2}$
are called \emph{operationally equivalent} and denoted by $M_{1}\sim M_{2}$
if they yield the same probability distribution of outcomes in every
preparation\footnote{Which are again identified by sets of laboratory instructions.}
of the system, that is if

\begin{eqnarray}
p(X|M_{1},P) & = & p(X|M_{2},P)\label{opeq}
\end{eqnarray}
Note that two operationally equivalent measurements are different
measurements if they are defined by different set of instructions.\footnote{Operational equivalence can be introduced into an operational theory
inductively and successively: one starts with a set of measurements
and preparations and render measurements equivalent which provide
the same outcome statistics in every preparations. This equivalence
class is relative to the set of preparations; a new preparation procedure
can break down operational equivalence if it discerns some measurements
with respect to their outcome statistics.}

A maximal set of basic measurements which can be performed simultaneously
on a system in an operational theory is called a \emph{context}. $M_{1}$
and $M_{2}$ are in the same context if and only if $M_{1}\wedge M_{2}\in\mathcal{M}$.
If $\{M_{1},M_{2},M_{3}\}$ is a context, then we call $M_{1}\wedge M_{2}\wedge M_{3}$
a \emph{maximally simultaneous measurement} and $M_{1}\wedge M_{2}$ a
\emph{non-maximally simultaneous measurement}. The set of all contexts
is a \emph{compatibility structure} of the theory. If in an operational
theory there are no two measurements which can be simultaneously measured,
then the compatibility structure is the empty set. We also refer to
such operational theories as \emph{trivial}. 

We call two operational theories \emph{operationally equivalent} (with
respect to their measurement) if any measurement in the one theory
is operationally equivalent with a measurement in the other theory
or with a marginalization thereof (and the preparations are the same). 

We call an operational theory \textit{non-disturbing}\footnote{Or \emph{no-signaling}, if the measurements are spacelike separated.}
if no conditional probability depends on whether the measurements
are performed alone or along with simultaneous measurements, that
is:
\begin{eqnarray}
p(X^{(i)}|M_{i},P)=p(X^{(i)}|M_{i}\wedge M_{j},P)\label{nondist}
\end{eqnarray}
for any simultaneous measurement $M_{i}\wedge M_{j}\in\mathcal{M}$; otherwise,
the operational theory is called \textit{disturbing}. Obviously, trivial
operational theories are non-disturbing. 

\begin{figure}[h]
\centerline{\resizebox{10cm}{!}{\includegraphics{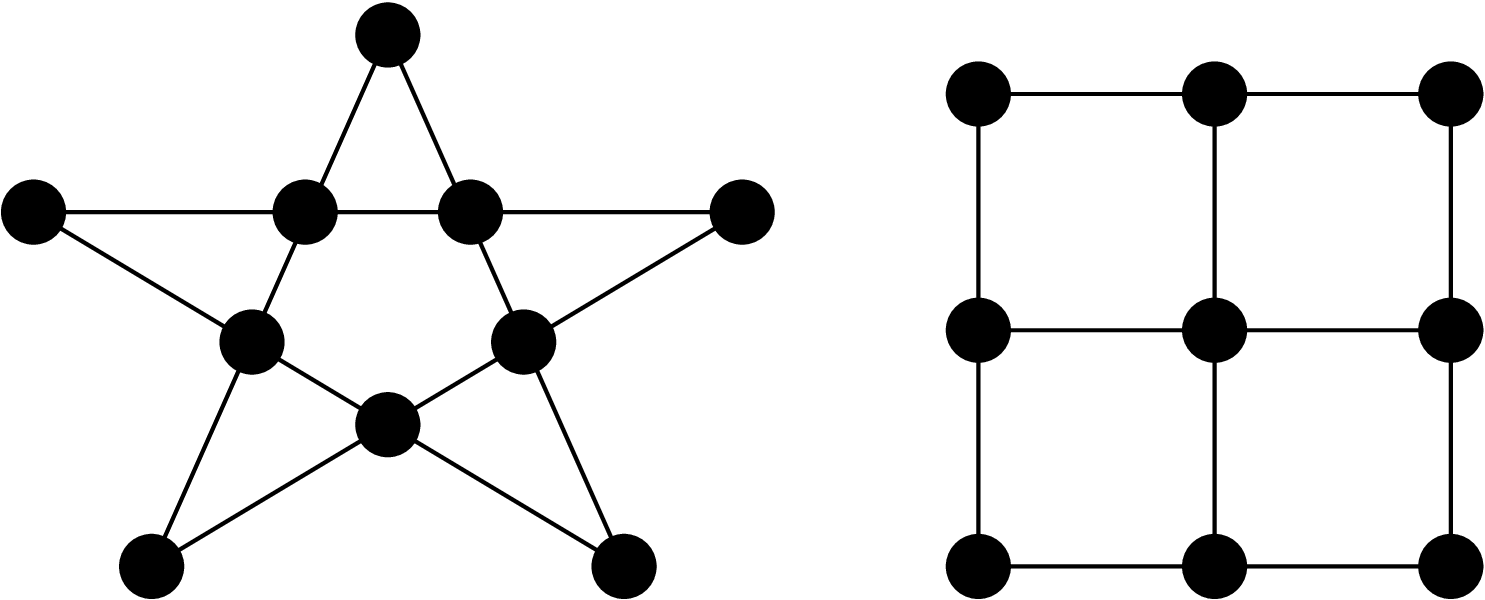}}}
\caption{The GHZ graph and Peres-Mermin graph}
\label{Fig:Bridge_graph1}
\end{figure}

\vspace{.1in} Next, we introduce a graph theoretical representation of operational
theories borrowed from the literature on the Kochen-Specker theorems
(Kochen and Specker, 1967). In Figure\ \ref{Fig:Bridge_graph1},
we depicted the graph of two Kochen-Specker theorems, the GHZ theorem
(Greenberger et al., 1990) on the left and the Peres-Mermin square
(Peres, 1990; Mermin, 1993) on the right. The vertices of the graph
represent self-adjoint operators and a subset of vertices is connected
by a (hyper)edge\footnote{A hyperedge can connect more than two vertices.}
if and only if the corresponding operators are pairwise commutating.
In the GHZ graph one has 10 operators and 5 commuting subsets; in
the Peres-Mermin graph one has 9 operators and 6 commuting subsets.
The (hyper)graph of most of the Kochen-Specker theorems is \emph{linear}
which means that each pair of hyperedges intersects in at most one
vertex.

In this paper, we take over this graphic representation and use it
in the framework of the operational theories but with a different
meaning. Vertices will represent here basic measurements and (hyper)edges
will represent maximal sets of simultaneous measurements, that is
contexts. In this interpretation, the above GHZ graph represents a
non-trivial operational theory with 10 basic measurement arranged
in 5 contexts and the Peres-Mermin graph represents a non-trivial
theory with 9 basic measurement and 6 contexts. The (hyper)graph of
both theory is linear: each basic measurement is featuring in exactly
two contexts.

\section{Ontological models}

\label{Sec:ontmod}

The role of an \textit{ontological model} (hidden variable model)
is to account for the conditional probabilities of an operational
theory in terms of underlying \textit{ontic states} (hidden variables,
elements of reality, beables, real states) of the measured system.
Let the set of ontic states be $\mathcal{\mathcal{L}\mathord{=}}\{\Lambda_{1},\Lambda_{2},\dots\}$
and let the random variable over $\mathcal{L}$ be $\Lambda.$ An
ontological model specifies a \textit{probability distribution} over
the ontic states associated with each preparation: 
\begin{equation}
p(\Lambda|P)\label{ont1}
\end{equation}
and a set of \textit{response functions} that is a set of conditional
probabilities associated with every measurement and every ontic state:
\begin{equation}
p(X|M,\Lambda)\label{ont2}
\end{equation}
again with the obvious normalizations. Assuming the independence of
the probability distributions from the measurements, called \textit{no-conspiracy}:
\begin{equation}
p(\Lambda|M,P)=p(\Lambda|P)\label{nocons}
\end{equation}
and the independence of the response functions from the preparations
in which the ontic states are featuring, called \textit{$\l$-sufficiency}:
\begin{equation}
p(X|M,P,\Lambda)=p(X|M,\Lambda)\label{lsuff}
\end{equation}
and using the theorem of total probability, one can recover the operational
theory from the ontological model in terms of the probability distributions
and response functions:

\begin{equation}
p(X|M,P)=\sum_{\Lambda}p(X|M,\Lambda)\,p(\Lambda|P)\label{recov}
\end{equation}

An ontological model is called \textit{outcome-deterministic (value-definite)}
if

\begin{equation}
p(X|M,\Lambda)\in\{0,1\}\label{determ}
\end{equation}
otherwise it is called \textit{outcome-indeterministic}.

\vspace{.1in} Next, we define two different and logically independent concepts
of noncontextuality (see Hofer-Szab\'o, 2021a, b, 2022).\emph{ }\textit{\emph{First,
an ontological model is called }}\textit{simultaneous noncontextual}\textit{\emph{
if every ontic state determines the probability of the outcomes of
every measurement independently of what other measurements are simultaneously
performed, that is}}\emph{ }
\begin{equation}
p(X^{(i)}|M_{i},\Lambda)=p(X^{(i)}|M_{i}\wedge M_{j},\Lambda)\label{NC1}
\end{equation}
for any simultaneous measurement $M_{i}\wedge M_{j}\in\mathcal{M}$\textit{\emph{;
otherwise the model is called }}\textit{simultaneous contextual.}\textit{\emph{
Simultaneous}}\emph{ }noncontextuality is a kind of inference to the
best explanation for why an operational theory is non-disturbing:
if the ontological model for an operational theory is noncontextual
in the sense of (\ref{NC1}), then---assuming no-conspiracy (\ref{nocons})
and $\l$-sufficiency (\ref{lsuff})---one can show that the operational
theory is non-disturbing (\ref{nondist}).

\textit{\emph{Second, an ontological model is called }}\textit{measurement
noncontextual}\textit{\emph{ if any two}}\textit{ operationally equivalent}\textit{\emph{
measurements, that is $M_{i},M_{j}\in M$}} \textit{\emph{which have
the same probability distribution of outcomes in every preparation}}
\begin{equation}
p(X|M_{i},P)=p(X|M_{j},P)\label{NC2a}
\end{equation}
\textit{\emph{also have the same probability distribution of outcomes
in every ontic state}}
\begin{equation}
p(X|M_{i},\Lambda)=p(X|M_{j},\Lambda)\label{NC2b}
\end{equation}
\textit{\emph{Otherwise the model is called }}\textit{measurement
contextual.} Measurement noncontextuality is again a kind of inference
to the best explanation; in this case the explanation of operational
equivalence: (\ref{NC2b})---together with no-conspiracy (\ref{nocons})
and $\l$-sufficiency (\ref{lsuff})---implies (\ref{NC2a}) (Hofer-Szab\'o,
2021a, Lin 2021).

In quantum mechanics where operationally equivalent measurements $M_{1}\sim M_{2}$
are represented by the same operator ${\bf O}$, measurement noncontextuality
is just the requirement that the response functions of an ontological
model should depend only on the operator and not on which specific
measurement is realizing the operator, that is
\[
p(X|M_{1},\Lambda)=p(X|M_{2},\Lambda)=p(X|{\bf O},\Lambda)
\]

Note, that trivial operational theories are trivially simultaneously
noncontextual (since there are no simultaneous measurements) but they
still can be measurement contextual. Also note that although simultaneous
noncontextuality and measurement noncontextuality are different and
logically independent notions, in case of non-disturbing theories
measurement noncontextuality implies simultaneous noncontextuality:
if $M_{j}$ does not disturb $M_{i}$, then (\ref{NC2a}) holds for
$M_{i}$ and $M_{i}\wedge M_{j}$ (with $X=X^{(i)}$), but then, due
to measurement noncontextuality, also (\ref{NC2b}), which is just
simultaneous noncontextuality (\ref{NC1}).

\section{Trivialization}

\label{Sec:trivialization}

With the framework of operational theories and ontological models
in hand, we can now formulate the main claim of our paper more precisely.
This claim was the following: ontological models for quantum mechanics,
and generally, for any operational theory sensitively depend on which
measurement we choose from the class of operationally equivalent measurements.
To show this dependence, I will investigate operational theories in
pairs such that the two theories have operationally equivalent measurements
but the first operational theory \emph{does have} and the second operational
theory \emph{does not have} simultaneous measurements. In other words,
the first theory has a non-trivial compatibility structure and the
second theory has a trivial one. More precisely, I will provide a
construction, which I call \emph{trivialization}, assigning to any
non-trivial operational theory a trivial theory. This construction
will yield us pairs of operational theories which then can be compared
with respect to the ontological models they admit and with respect
to such properties as the causal structure, contextuality, fine-tuning,
etc. We will see how sensitively the ontological models depend on
whether the operational model is trivial or not. 

In this section, I will only outline the procedure of trivialization
and show some of its graph theoretical properties. In the next two
sections, I will compare---on a classical, a quantum, and a super-quantum
mechanical example---the ontological models of three non-trivial
and three corresponding trivial theories. In Section \ref{Sec:QMech},
I return to the quantum mechanical example in order to highlight that
operational theories realizing the same set of operators by different
measurements can be vastly different. 

Let us now turn to the trivialization. Consider an operational theory
with a non-trivial compatibility structure that is a theory which
comprises both basic and simultaneous measurements. \emph{Trivialization}
then consists in the following procedure:

\vspace{.1in} \emph{Replace some (or perhaps all) of the measurements in the
non-trivial operational theory with new, operationally equivalent
measurements such that in the resulting operational theory there are
no two measurements which can be performed }simultaneously\emph{. }

\vspace{.1in} An example might help. Consider a non-disturbing operational
theory with the following set of measurements:
\[
\mathcal{M_{\mbox{}}=}\{M_{1},M_{2},M_{3},M_{4},M_{5},\,M_{1}\wedge M_{2},\,M_{1}\wedge M_{2}\wedge M_{3},\,M_{1}\wedge M_{4}\}
\]
The theory has five basic measurements, $M_{1},M_{2},M_{3},M_{4},M_{5}$;
one non-maximal simultaneous measurement $M_{1}\wedge M_{2}$; and
two maximally simultaneous measurements, $M_{1}\wedge M_{2}\wedge M_{3}$
and $M_{1}\wedge M_{4}$ corresponding to the contexts $\{M_{1},M_{2},M_{3}\}$
and $\{M_{1},M_{4}\}$, respectively. 

The trivialization of $\mathcal{M}$ consist in the the following
steps. We keep the basic measurement $M_{5}$ and replace $M_{1}\wedge M_{2}\wedge M_{3}$
and $M_{1}\wedge M_{4}$ with two new measurements $M_{123}$ and
$M_{14}$ with outcome sets $\mathcal{X}^{(123)}$ and $\mathcal{X}^{(14)}$,
respectively. Note that $M_{123}$ and $M_{14}$ are completely new
measurements and not the conjunctions $M_{1}\wedge M_{2}\wedge M_{3}$
and $M_{1}\wedge M_{4}$. The simple reason we denote them by multiple
indices is to relate them to these conjunctions. These new measurements
are operationally equivalent to the old maximally simultaneous measurements:\footnote{Obviously, it is a theoretical-experimental question whether such
new measurements exist. In the quantum mechanical scenario which we
use in the paper they do, and we will explicitly show them in Section
\ref{Sec:QMech}.}

\[
M_{123}\sim M_{1}\wedge M_{2}\wedge M_{3}\quad\mbox{and}\quad M{}_{14}\sim M_{1}\wedge M_{4}
\]
This means that for every preparation:
\begin{eqnarray*}
p(f(X^{(1)}\wedge X^{(2)}\wedge X^{(3)})|M_{123},P) & = & p(X^{(1)}\wedge X^{(2)}\wedge X^{(3)}|M_{1}\wedge M_{2}\wedge M_{3},P)\\
p(g(X^{(1)}\wedge X^{(4)})|M_{14},P) & = & p(X^{(1)}\wedge X^{(4)}|M_{1}\wedge M_{4},P)
\end{eqnarray*}
where $f$ is a bijection mapping the outcomes $\mathcal{X}^{(1)}\times\mathcal{X}^{(2)}\times\mathcal{X}^{(3)}$
of $M_{1}\wedge M_{2}\wedge M_{3}$ to the outcomes $\mathcal{X}^{(123)}$
of $M_{123}$, and $g$ is another bijection mapping the outcomes
$\mathcal{X}^{(1)}\times\mathcal{X}^{(4)}$ of $M_{1}\wedge M_{4}$
to the outcomes $\mathcal{X}^{(14)}$ of $M_{14}$. 

The measurements $M_{5},M_{123}$ and $M_{14}$ are basic measurements
and the operational theory is \emph{trivial} since no measurements
can be simultaneously measured. Since the theory is non-disturbing,
the other five old measurements, $M_{1},M_{2},M_{3},M_{4}$ and $M_{1}\wedge M_{2}$
will operationally equivalent to certain coarse-graining or marginalization
of the new basic measurements, $M_{123}$ and $M_{14}$. To see this,
let us introduce the following notation. Let $f(X^{(1)})$ denote
the \emph{union} of those outcomes of $M_{1}\wedge M_{2}\wedge M_{3}$
which are assigned to the outcome $X^{(1)}$ of $M_{1}$ by the bijection
$f$. Let $f(X^{(2)})$ and $f(X^{(3)})$ be defined in a similar
way. Furthermore, let $f(X^{(1)}\wedge X^{(2)})$ denote the \emph{union}
of those outcomes of $M_{1}\wedge M_{2}\wedge M_{3}$ which are assigned
to the outcome $X^{(1)}\wedge X^{(2)}$ of $M_{1}\wedge M_{2}$ by
the bijection $f$. Finally, let $g(X^{(1)})$ denote the \emph{union}
of those outcomes of $M_{1}\wedge M_{4}$ which are assigned to the
outcome $X^{(1)}$ of $M_{1}$ and let $g(X^{(4)})$ be similarly
defined. Then, the five old measurements will be operationally equivalent
to the following marginalization of the new basic measurements:

\begin{eqnarray*}
p(f(X^{(1)})|M_{123},P)=p(g(X^{(1))}|M_{14},P) & = & p(X^{(1)}|M_{1},P)\\
p(f(X^{(2)})|M_{123},P) & = & p(X^{(2)}|M_{2},P)\\
p(f(X^{(3)})|M_{123},P) & = & p(X^{(3)}|M_{3},P)\\
p(g(X^{(4))}|M_{14},P) & = & p(X^{(4)}|M_{4},P)\\
p(f(X^{(1)}\wedge X^{(2)})|M_{123},P) & = & p(X^{(1)}\wedge X^{(2)}|M_{1}\wedge M_{2},P)
\end{eqnarray*}
For the operational equivalence of these marginalizations, we will
use the short hand

\[
M_{123}^{(1)}\sim M_{14}^{(1)}\sim M_{1},\quad M_{123}^{(2)}\sim M_{2},\quad M_{123}^{(3)}\sim M_{3},\quad M_{14}^{(4)}\sim M_{4},\quad M_{123}^{(12)}\sim M_{1}\wedge M_{2}
\]
where $M_{123}^{(1)}$ denotes the measurement that we first perform
the measurement $M_{123}$, and then  coarse-grain the outcomes into
blocks such that each block corresponds---due to the bijection $f$---
to an outcome $X^{(1)}$ of $M_{1}$.

To sum up, the new operational theory will be the following 

\[
\mathcal{M}'=\{M_{123},M_{14},M_{5}\}
\]
The two operational theories $\mathcal{M}$ and $\mathcal{M}'$ are
operationally equivalent, $\mathcal{M}\mathcal{\,\sim M}'$, since
any measurement in $\mathcal{M}$ is operationally equivalent to a
measurement or a specific marginalization of a measurement in $\mathcal{M}'$
and vica versa. They are, however, different since they contain different
measurements (except for the common $M_{5}$). $\mathcal{M}$ is non-trivial
but $\mathcal{M}'$ is trivial, it contains no simultaneous measurements. 

Note again that even though $M_{14}$ is indexed by two indices, it
is just as a basic measurement in the new operational theory as $M_{1}$
and $M_{4}$ was in the old theory. The only reason why we use these
multiple indices is to be able to relate $M_{14}$ to $M_{1}$ and
$M_{4}$ simply by marginalization and operational equivalence. This
 notation, however, should not blur the fact that $M_{14}$ can be
a simple measurement.\footnote{Just to stress this fact again, one need not think of $M_{14}$ as
being two measurements performed on two different subsystems, as in
the usual spin measurement scenarios in quantum mechanics. $M_{14}$
can also be a measurement on a localized system. }

\begin{figure}[h]
\centerline{\resizebox{14cm}{!}{\includegraphics{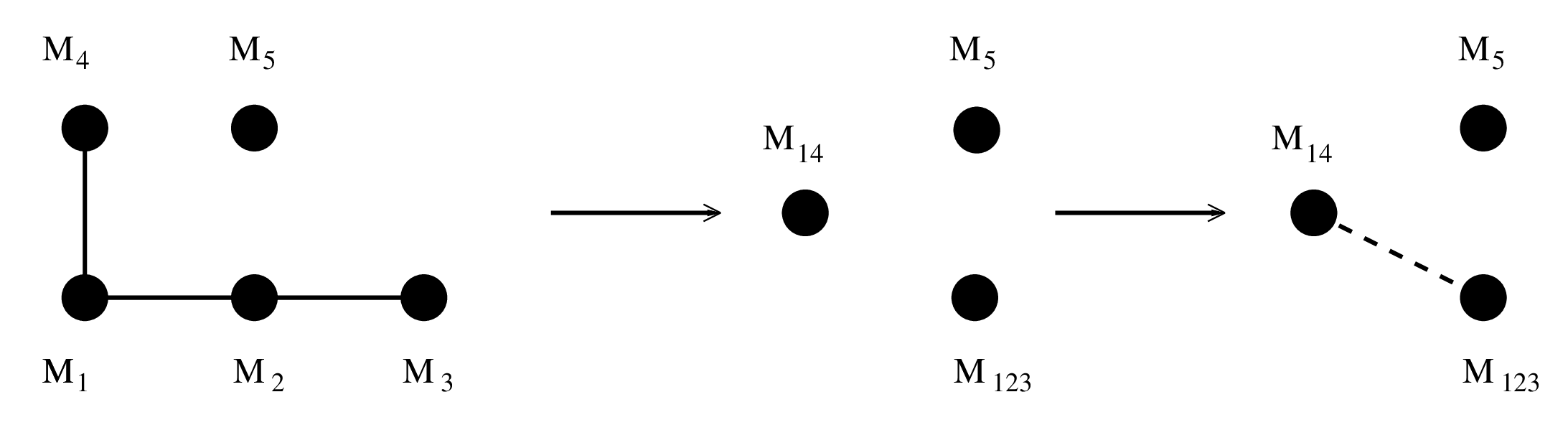}}}
\caption{The graph and line graph of the toy operational theory}
\label{Fig:Bridge_graph22}
\end{figure}

\vspace{.1in} Now, let us turn to the graphic representation of the trivialization.
As we saw, trivialization results in an operational theory with trivial
compatibility structure. If we represent this new operational theory
by a graph, this graph will have only vertices but no edges. The first
two graphs in Figure\ \ref{Fig:Bridge_graph22} show the graph of
our above mini operational theory and the trivialized new theory.

We could stop at this point but then some information would be lost,
namely, that certain marginalizations of the new measurements are
operationally equivalent. To preserve this information, we add (hyper)edges
to the graph of the new theory \emph{with the following meaning}:
we draw an (hyper)edge between a set of vertices in the trivial theory,
if the corresponding basic measurements have operationally equivalent
marginalizations.\footnote{Or equivalently, if the contexts conforming to the maximally simultaneous
measurements in the non-trivial theory (which are represented by the
vertices in the trivial theory) had at least one common basic measurement.} For example, the graph of our mini operational will get an edge (see
the third graph in Figure\ \ref{Fig:Bridge_graph22}) because the
appropriate marginalization $M_{123}$ and $M_{14}$ are operationally
equivalent to one another (both being operationally equivalent to
$M_{1}$). Note that the (hyper)edges in the graph of the non-trivial
and trivial theories mean different things: in the non-trivial operational
theory they meant simultaneous measurability, while in the trivial
operational theory they mean having operationally equivalent marginalizations.
To express this difference, we use continuous lines in the non-trivial
operational theories and broken lines in the trivial ones.

\begin{figure}[h]
\centerline{\resizebox{11cm}{!}{\includegraphics{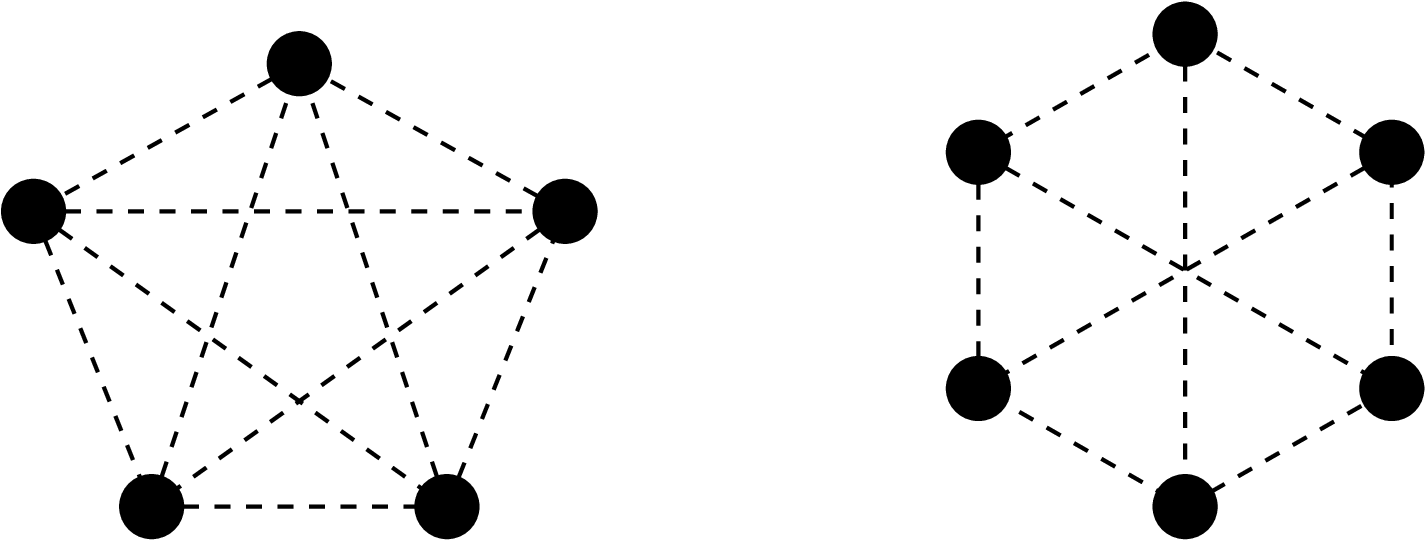}}}
\caption{The line graphs of the GHZ graph and Peres-Mermin graph}
\label{Fig:Bridge_linegraph1}
\end{figure}

This construction can be nicely represented graph theoretically by
simply taking the line graphs of the (hyper)graph of the non-trivial
operational theory. A \emph{line graph} $L(G)$ is constructed from
a graph $G$ such that for each (hyper)edge in $G$ we make a vertex
in $L(G)$ and for every two (hyper)edges in $G$ that have a vertex
in common, we make an edge between their corresponding vertices in
$L(G)$. The line graphs of the GHZ graph and Peres-Mermin graph,
for example, are depicted in Figure\ \ref{Fig:Bridge_linegraph1}.
The number of the vertices and edges flip in both line graphs: the
line graph of the GHZ graph contains 5 vertices and 10 edges, the
line graph of the Peres-Mermin graph contains 6 vertices and 9 edges.
Since both the GHZ graph and Peres-Mermin graph are linear, their
line graphs contain only edges but no hyperedges.

To sum up, in the graph $G$ of the non-trivial operational theory,
vertices represent the old basic measurements and (hyper)edges represented
contexts that is sets of simultaneous measurements. In the line graph
$L(G)$ of the trivialized theory, vertices represent the new basic
measurements but---since there are no simultaneous measurements---
the (hyper)edges mean something else: they connect vertices representing
measurements which have operationally equivalent marginalizations. 

\section{Three operational theories with non-trivial compatibility structure}

\label{Sec:nontrivi}

\begin{figure}[h]
\centerline{\resizebox{7cm}{!}{\includegraphics{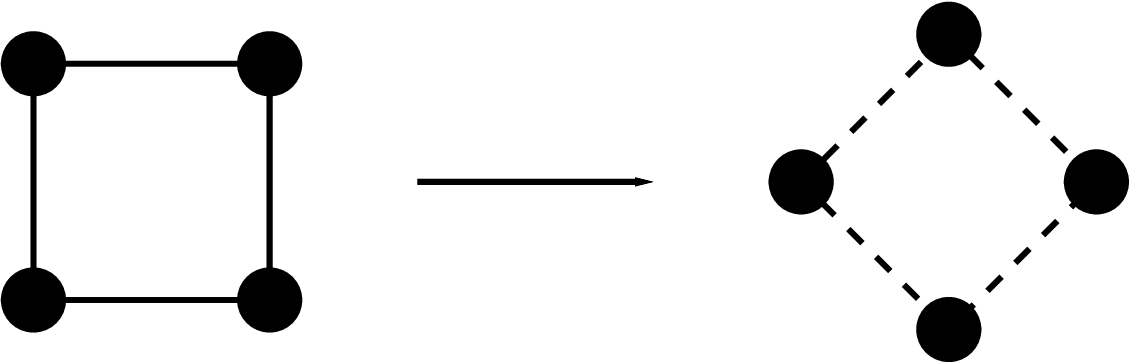}}}
\caption{The graph and line graph of the three operational theories}
\label{Fig:Bridge_graph3}
\end{figure}

In this Section, we consider three non-disturbing and non-trivial
operational theories, all of the form

\[
\mathcal{M_{\mbox{}}=}\{A_{0},A_{1},B_{0},B_{1},\,A_{0}\wedge B_{0},\,A_{0}\wedge B_{1},\,A_{1}\wedge B_{0},\,A_{1}\wedge B_{1}\}
\]
Each theory has the same four basic measurements $A_{0},A_{1},B_{0},B_{1}$
such that $A_{0},A_{1}$ have binary outcomes $X_{0},X_{1}$ and $B_{0},B_{1}$
have binary outcomes $Y_{0},Y_{1}$. In all three theories, any $A$-measurement
is simultaneously measurable with any $B$-measurement but neither
the two $A$-measurements nor the two $B$-measurements can be simultaneously
measured. In short, the compatibility structure of all three theories
will be
\[
\big\{\{A_{0},B_{0}\},\{A_{0},B_{1}\},\{A_{1},B_{0}\},\{A_{1},B_{1}\}\big\}
\]
Consequently, the graph (and line graph, see next section) depicted
in Figure\ \ref{Fig:Bridge_graph3} is the same for all three operational
theories. Since the graph is linear, the line graph contains only
edges and no hyperedges.

Let $A$ be a random variable over the measurements $\{A_{0},A_{1}\}$
and $B$ a random variable over the measurements over the measurements
$\{B_{0},B_{1}\}$. Similarly, let $X$ and $Y$ be random variables
over the outcomes $\{X_{0},X_{1}\}$ and $\{Y_{0},Y_{1}\}$, respectively,
such that $A,B,X,Y=0,1$. The operational theories differ in the preparations.
Each theory has only one preparation: the first one $P_{\tiny\mbox{CL}}$,
the second $P_{\tiny\mbox{EPR}}$, and the third $P_{\tiny\mbox{PR}}$.
We refer to the operational theories as a \emph{classical operational
theory}, the \emph{EPR-Bell situation}, and the \emph{Popescu-Rorhlich
(PR) box} (Popescu and Rohrlich, 1994), respectively.

The three operational theories can be characterized by the following
conditional probabilities:

\begin{eqnarray}
p(X|A,P)=p(Y|B,P) & = & \frac{1}{2}\label{OP}\\
p(X,Y|A,B,P_{\tiny\mbox{CL}}) & = & \begin{cases}
\begin{array}{cc}
\frac{1}{2} & \quad\mbox{if}\,\,X\oplus Y=0\\
0 & \mbox{otherwise}
\end{array}\end{cases}\label{CL}\\
p(X,Y|A,B,P_{\tiny\mbox{EPR}}) & = & \begin{cases}
\begin{array}{cc}
\frac{3}{8} & \quad\mbox{if}\,\,X\oplus Y=0\,\,\mbox{and}\,\,A\cdot B=0\\
\frac{1}{8} & \quad\mbox{if}\,\,X\oplus Y=1\,\,\mbox{and}\,\,A\cdot B=0\\
\frac{1}{2} & \quad\mbox{if}\,\,X\oplus Y=0\,\,\mbox{and}\,\,A\cdot B=1\\
0 & \quad\mbox{if}\,\,X\oplus Y=1\,\,\mbox{and}\,\,A\cdot B=1
\end{array}\end{cases}\label{EPR}\\
p(X,Y|A,B,P_{\tiny\mbox{PR}}) & = & \begin{cases}
\begin{array}{cc}
\frac{1}{2} & \quad\mbox{if}\,\,X\oplus Y=A\cdot B\\
0 & \mbox{otherwise}
\end{array}\end{cases}\label{PR}
\end{eqnarray}
where $P$ is a variable over $\mathcal{P}=\{P_{\tiny\mbox{CL}},P_{\tiny\mbox{EPR}},P_{\tiny\mbox{PR}}\}$
and $\oplus$ is the sum modulo 2. We come back to the quantum mechanical
representation of the EPR-Bell situation in Section \ref{Sec:QMech}.

\textit{\emph{All three operational theories are non-disturbing: }}

\textit{\emph{
\begin{eqnarray}
p(X|A,P) & = & p(X|A,B,P)=\frac{1}{2}\label{nondist1}\\
p(Y|B,P) & = & p(Y|A,B,P)=\frac{1}{2}\label{nondist2}
\end{eqnarray}
}}but the\textit{\emph{y they represent three different classes of
theories. The first is a classical theory, the second is a quantum
mechanical, the third is a super-quantum mechanical theory. This difference
is manifested in the satisfaction/violation of the}} CHSH inequality
(Clauser, Horne, Shimony, and Holt, 1969). Namely, the CHSH expression

\begin{equation}
\mbox{CHSH}_{P}=\left\langle A_{0},B_{0}\right\rangle _{P}+\left\langle A_{0},B_{1}\right\rangle _{P}+\left\langle A_{1},B_{0}\right\rangle _{P}-\left\langle A_{1},B_{1}\right\rangle _{P}\label{CHSH}
\end{equation}
\textit{\emph{where 
\[
\left\langle A,B\right\rangle _{P}=p(X\oplus Y=0|A,B,P)-p(X\oplus Y=1|A,B,P)
\]
is $2$ for the classical theory, satisfying the CHSH inequality,
$|\mbox{CHSH}_{P}|\leqslant2$; it is $2.5$ for the EPR-Bell situation,
violating (not maximally) the CHSH inequality; and $4$ for the PR
box which is beyond the Tsirelson bound $2\sqrt{2}$. }}

\begin{figure}[h]
\centerline{\resizebox{16cm}{!}{\includegraphics{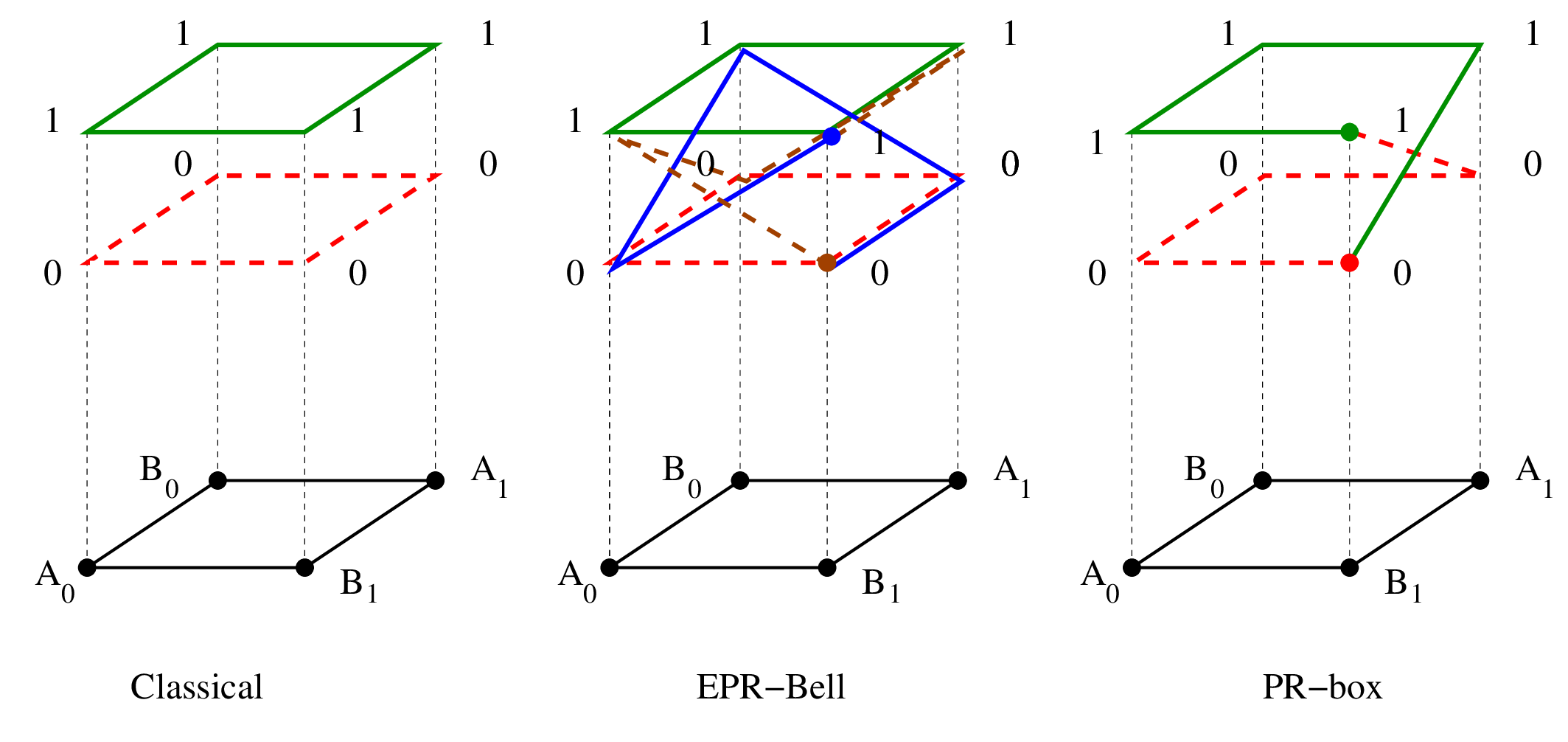}}} \caption{Bundle diagrams of the ontological models for the operational theories
with non-trivial compatibility structure}
\label{Fig:Bridge1}
\end{figure}

\vspace{.1in} \textit{\emph{Next, we construct an ontological model for each
operational theory. The exact probabilistic specification of the models
in terms of distributions and response functions is given in the Appendix.
From our perspective, however, it }}will be more instructive\textit{\emph{
to look at the }}\emph{bundle diagrams}\textit{\emph{ (see Abramsky
et al., 2017; Abramsky \& Brandenburger, 2011)}} of the models depicted
in Figure\ \ref{Fig:Bridge1}. 

\textit{\emph{First, look at the ``cuboid'' of the classical model
on the left. The quadrangle at the bottom is the base space of the
bundle, actually the graph of the operational theory ``laid down''.
It consists of four vertices representing the four measurements }}$A_{0},A_{1},B_{0},B_{1}$
such that two measurements are connected if and only if they are in
the same context.\textit{\emph{ The vertical broken lines are the
fibers of the bundle. The two vertices on a given fiber at different
heights denoted by $0$ and $1$ represent the outcomes of the corresponding
measurements: $X=0,1$ for $A=0,1$ and $Y=0,1$ for $B=0,1$. Now,
there are two quadrangles in the figure, one connecting the upper
vertices of the adjacent fibers and one connecting the lower vertices.
Each quadrangle represents the response functions of the model for
a given ontic state. The green and continuous upper quadrangle represents
the ontic state $\Lambda_{1}$. In this ontic state the outcome of
each measurement is 1. (The outcome of }}$A_{0},A_{1}$ is $X_{1}$
and the outcome of $B_{0},B_{1}$ is $Y_{1}.)$ \textit{\emph{The
red and broken lower quadrangle represents the ontic state $\Lambda_{0}$
for which each outcome is 0. The model is outcome-deterministic. It
also fixes the outcomes of the simultaneous measurements in the different
contexts such that no outcome of any measurement in any ontic state
depends on whether a simultaneous measurement is also performed. Thus,
the model is simultaneous noncontextual. Moreover, the model is also
measurement noncontextual: in both ontic state the outcome of any
two operationally equivalent measurements is the same. }}Setting the
probability of both ontic states to $\frac{1}{2}$, the operational
theory can be recovered. 

Let us now go over to the\textit{\emph{ bundle diagram of the PR-box
on the right of }}Figure\ \ref{Fig:Bridge1}\textit{\emph{. Again,
we have two ontic states $\Lambda_{1}$ and $\Lambda_{0}$ but }}the
green and red lines do not close up now. They are discontinuous at
the fibre of $B_{1}$. To avoid ambiguity with respect to the outcome
of $B_{1}$, we put a dot at the one end of both discontinuous lines.
This dot indicates the outcome of $B_{1}$ if measured \emph{alone}
and not together with $A_{0}$ or $A_{1}$ (when the outcome of $B_{1}$
is indicated by the value of the appropriate segment of the green
or red lines connecting the fibre of $B_{1}$ with the fibre of $A_{0}$
or $A_{1}$). Thus, the model is outcome deterministic. However, it
is simultaneous contextual: 
\begin{equation}
\delta_{Y,\Lambda}=p(Y|B_{1},\Lambda)\neq p(Y|A_{1}\wedge B_{1},\Lambda)=\delta_{Y\oplus1,\Lambda}\label{NC1-PR}
\end{equation}
That is performing the measurement $B_{1}$ in the ``green'' ontic
state, \textit{\emph{$\Lambda_{1}$,}} together with $A_{1}$, the
outcome of $B_{1}$ will be $Y_{0}$, while performing $B_{1}$ together
with $A_{0}$, the outcome will be $Y_{1}$; and \emph{vice versa}
for the ``red'' ontic state, \textit{\emph{$\Lambda_{0}$}}. Since
simultaneous contextuality implies measurement contextuality for non-disturbing
theories, the model for the PR-box will also be measurement contextual.
Indeed, 

\begin{equation}
p(Y|B_{1},P_{\tiny\mbox{PR}})=p(X|A_{1}\wedge B_{1},P_{\tiny\mbox{PR}})\label{NC2a-PR}
\end{equation}
\textit{\emph{despite the fact that inequality }}(\ref{NC1-PR}) holds.
We can recover the PR-box theory again by setting the probability
of both ontic states to $\frac{1}{2}$.

Finally, the\textit{\emph{ bundle diagram in the middle of }}Figure\ \ref{Fig:Bridge1}
represents an ontological model for the EPR-Bell scenario. Here we
have four ontic states portrayed by lines of different color and style.
The ``green'' and ``red'' ontic states are outcome deterministic
and noncontextual in both senses. The ``blue'' and ``brown'' ontic
states, however, are outcome deterministic but simultaneous and hence
measurement contextual: their lines do not close on the fibre of $B_{1}$.
This means that in these ontic states the outcome of $B_{1}$ will
be different when measured alone and when co-measured with $A_{0}$
or $A_{1}$. The dots at one end of the lines indicate the outcomes
the outcome of $B_{1}$ when measured alone. By setting the probability
of the two noncontextual ontic states to $\frac{3}{8}$ and the probability
of the two contextual ontic states to $\frac{1}{8}$, the probabilities
of the EPR-Bell scenario can be recovered (see Appendix).

To sum up, we constructed three (among the many) outcome-deterministic
ontological models for the three operational theories such that the
model for the classical theory is noncontextual (in both senses) and
the models for other two theories are contextual (again, in both senses).
This is in tune with the satisfaction and violation of the CHSH inequality
for the different theories.

\section{Three operational theories with trivial compatibility structure}

\label{Sec:trivi}

The three operational theories in the previous Section were non-trivial,
they had a non-trivial compatibility structure. Let us now ``trivialize''
them in the way outlined in Section \ref{Sec:trivialization} and
investigate the ontological models for these trivialized theories.
Trivialization consists in replacing each simultaneous measurement
\[
A_{0}\wedge B_{0},\quad A_{0}\wedge B_{1},\quad A_{1}\wedge B_{0},\quad A_{1}\wedge B_{1}
\]
with an operationally equivalent new basic measurement:

\begin{eqnarray*}
C_{00}\sim A_{0}\wedge B_{0}, & \quad & C_{01}\sim A_{0}\wedge B_{1}\\
C_{10}\sim A_{1}\wedge B_{0}, & \quad & C_{11}\sim A_{1}\wedge B_{1}
\end{eqnarray*}
Note that $C_{00},C_{01},C_{10}$ and $C_{11}$ cannot be measured
simultaneously.

Let the outcome space of each of the measurements in $\mathcal{C=}\big\{ C_{00},C_{01},C_{10},C_{11}\big\}$
be the same $\mathcal{Z}\mathcal{=}\big\{ Z_{00},Z_{01},Z_{10},Z_{11}\big\}$.
Let $C$ be a random variable over $\mathcal{C}$ assigning to every
measurement its index pair. Similarly, let $Z$ be a random variable
over $\mathcal{Z}$ assigning to every outcome its index pair. Both
$C$ and $Z$ can be expressed as a Cartesian product: $C=C_{1}\times C_{2}$
and $Z=Z_{1}\times Z_{2}$ where $C_{1}$ and $Z_{1}$ assign to every
measurement or outcome its first index and $C_{2}$ and $Z_{2}$ assign
the second index.

Since the old operational theory is non-disturbing, the following
marginalizations of the new basic measurements are operationally equivalent
to the old basic measurements:\footnote{Note that $Z_{1}$ is the \emph{union} of those two outcomes of $C_{00},C_{01},C_{10}$
and $C_{11}$ which has the same first index. In the terminology introduced
in Section \ref{Sec:trivialization}, $Z_{1}$ is the \emph{union}
of those outcomes which are assigned to the outcome $X$ of $A_{0}$
by a bijection $f_{0}$ or to the outcome $X$ of $A_{1}$ by a bijection
$f_{1}$ and also to the outcome $Y$ of $B_{0}$ by a bijection $g_{0}$
or to the outcome $Y$ of $B_{1}$ by a bijection $g_{1}$. To keep
the notation simple, we drop these bijections and write simply $Z_{1}$
instead of $f_{0}(X)$, $f_{1}(X)$, $g_{0}(Y)$ and $g_{1}(Y)$. }

\begin{eqnarray*}
p(Z_{1}|C_{00},P)=p(Z_{1}|C_{01},P) & = & p(X|A_{0},P)\\
p(Z_{1}|C_{10},P)=p(Z_{1}|C_{11},P) & = & p(X|A_{1},P)\\
p(Z_{2}|C_{00},P)=p(Z_{2}|C_{01},P) & = & p(Y|B_{0},P)\\
p(Z_{2}|C_{01},P)=p(Z_{2}|C_{11},P) & = & p(Y|B_{1},P)
\end{eqnarray*}
or using the short hand introduced in Section \ref{Sec:trivialization}:
\begin{eqnarray*}
C_{00}^{(1)}\sim C_{01}^{(1)}\sim A_{0}, & \quad & C_{10}^{(1)}\sim C_{11}^{(1)}\sim A_{1}\\
C_{00}^{(2)}\sim C_{10}^{(2)}\sim B_{0}, & \quad & C_{01}^{(2)}\sim C_{11}^{(2)}\sim B_{1}
\end{eqnarray*}
That is, the old basic measurements $A_{0},A_{1},B_{0},B_{1}$ can
be recovered as marginalizations of the new basic measurements. Thus,
the new trivialized operational theories will be:

\[
\mathcal{M}'=\{C_{00},C_{01},C_{10},C_{11}\}
\]
$\mathcal{M}'$ is operationally equivalent to $\mathcal{M}$ introduced
in the previous section, $\mathcal{M}\mathcal{\,\sim M}'$, or more
precisely: 
\[
\mathcal{M}_{CL}\,\mathcal{\sim M}'_{CL},\quad\mathcal{\quad M}_{EPR}\mathcal{\,\sim M}'_{EPR},\quad\quad\mathcal{M}_{PR}\mathcal{\,\sim M}'_{PR}
\]
The line graph of $\mathcal{M}$' is depicted on the right side of
Figure\ \ref{Fig:Bridge_graph3}. 

The three trivial operational theories can be characterized by the
following conditional probabilities:

\begin{eqnarray}
p(Z|C,P_{\tiny\mbox{CL}}) & = & \begin{cases}
\begin{array}{cc}
\frac{1}{2} & \quad\mbox{if}\,\,Z_{1}\oplus Z_{2}=0\\
0 & \mbox{otherwise}
\end{array}\end{cases}\label{CL-1}\\
p(Z|C,P_{\tiny\mbox{EPR}}) & = & \begin{cases}
\begin{array}{cc}
\frac{3}{8} & \quad\mbox{if}\,\,Z_{1}\oplus Z_{2}=0\,\,\mbox{and}\,\,C_{1}\cdot C_{2}=0\\
\frac{1}{8} & \quad\mbox{if}\,\,Z_{1}\oplus Z_{2}=1\,\,\mbox{and}\,\,C_{1}\cdot C_{2}=0\\
\frac{1}{2} & \quad\mbox{if}\,\,Z_{1}\oplus Z_{2}=0\,\,\mbox{and}\,\,C_{1}\cdot C_{2}=1\\
0 & \quad\mbox{if}\,\,Z_{1}\oplus Z_{2}=1\,\,\mbox{and}\,\,C_{1}\cdot C_{2}=1
\end{array}\end{cases}\label{EPR-1}\\
p(Z|C,P_{\tiny\mbox{PR}}) & = & \begin{cases}
\begin{array}{cc}
\frac{1}{2} & \quad\mbox{if}\,\,Z_{1}\oplus Z_{2}=C_{1}\cdot C_{2}\\
0 & \mbox{otherwise}
\end{array}\end{cases}\label{PR-1}
\end{eqnarray}

Observe that the probabilistic description of the trivial operational
theories is formally analogous with that of the non-trivial theories
of the previous section: we obtain equations (\ref{CL})-(\ref{PR})
from (\ref{CL-1})-(\ref{PR-1}) by simply replacing $C_{1}$, $C_{2}$,
$Z_{1}$, $Z_{2}$ with $A$, $B$, $X$, $Y$, respectively. The
measurements and outcomes, however, are different in the two theories.

\textit{\emph{All three operational theories are non-disturbing in
a trivial sense: there are no simultaneous measurements. Therefore,
the CHSH inequalities cannot be defined. Again, one can construct
an ontological model for each operational theory. The distribution
of ontic states is the same as in the models for the non-trivial theories.
The response functions are obtained from those of the non-trivial
theory by}} simply replacing $A$, $B$, $X$, $Y$ with $C_{1}$,
$C_{2}$, $Z_{1}$, $Z_{2}$. All this is specified\textit{\emph{
in the Appendix and visualized in }}Figure\ \ref{Fig:Bridge1v}.
\begin{figure}[h]
\centerline{\resizebox{16cm}{!}{\includegraphics{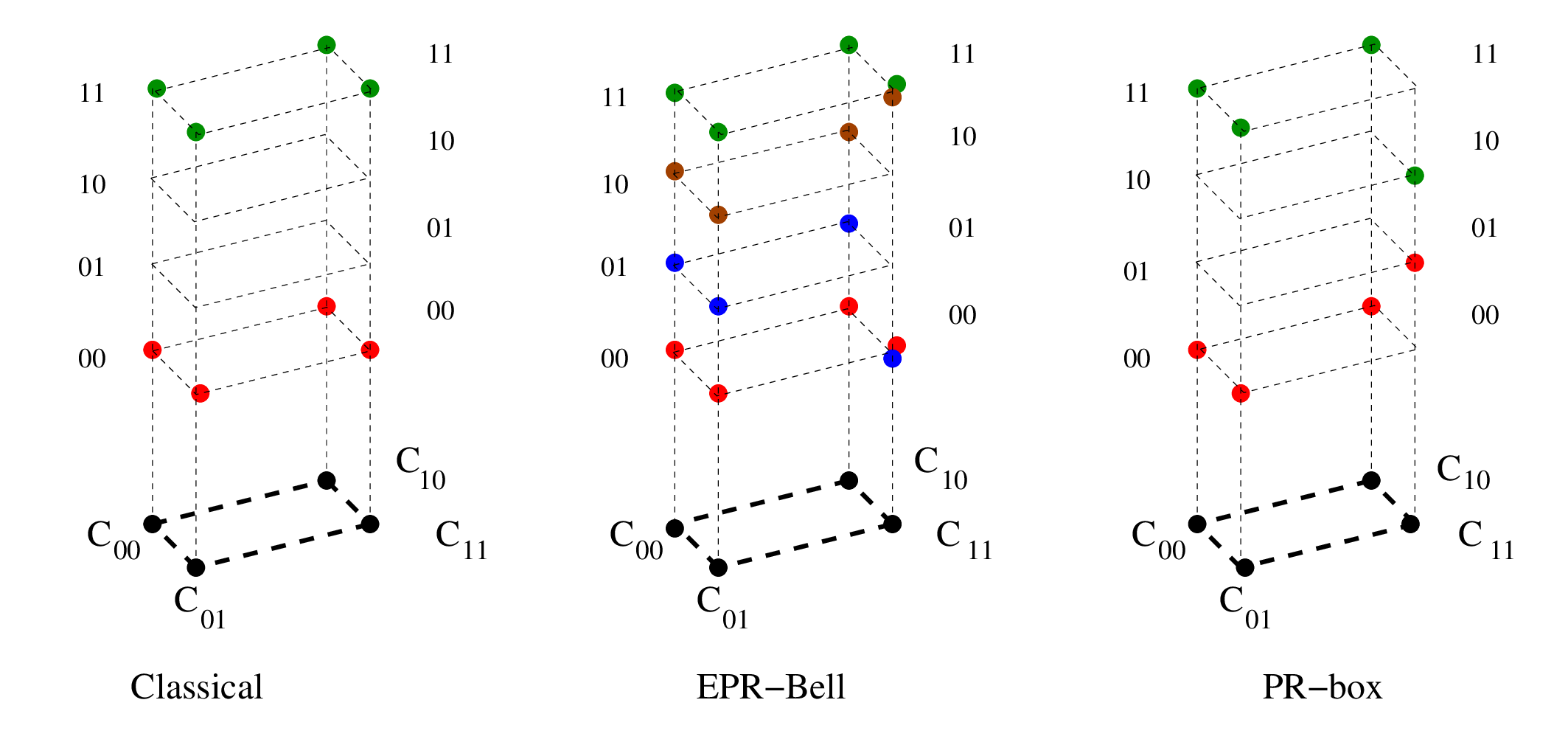}}} \caption{Bundle diagrams of the ontological models for the operational theories
with trivial compatibility structure}
\label{Fig:Bridge1v}
\end{figure}
As can be seen, the lines representing the outcomes of simultaneous
measurements have disappeared. Each basic measurement has a definite
outcome in every ontic state denoted by a dot at the appropriate height
on the fibre corresponding to the measurement. In the classical theory
and in the PR box there are two ontic states (``green'' and ``red''),
in the EPR-Bell scenario there are four ontic states (``green'',
``red'', ``blue'' and ``brown''). All three models are outcome-deterministic
and simultaneously non-contextual since there are no simultaneous
measurement. But the non-classical (EPR and PR) models are measurement
contextual. Certain marginalizations of the measurements, for example,
$C_{01}^{(2)}$ and $C_{11}^{(2)}$ are operationally equivalent.
Still, both the ``blue'' and ``brown'' ontic states in the EPR
model and ``green'' and ``red'' ontic states in the PR model assign
different outcomes to them. This shows that measurement noncontextuality
is a stronger concept than simultaneous noncontextuality. 

\section{The causal structure of the ontological models}

\label{Sec:causal}

\textit{\emph{Let us turn now to the causal structure of the ontological
models. Since these models provide information only about the probabilistic
relations of the events and not about their spatiotemporal or other
relations, the reconstruction of the causal structure will rely solely
on these probabilistic information. The machinery to deduce causal
relations from probabilistic relations is known as }}\textit{causal
discovery algorithms}\textit{\emph{ and was introduced in (Pearl,
2009; Spirtes, Glymour, Scheines, 2001). These algorithms do not make
use of the full probabilistic setting, they use only the conditional
and unconditional independence relations to construct a causal graph.
A causal graph is a directed acyclic graph (DAG),}}\footnote{\textit{\emph{Note that these causal graphs are different from the
graphs and line graphs used in the previous Sections representing
compatibility structure and common marginalization.}}}\textit{\emph{ where the vertices represent random variables and the
directed edges represent causal relevance between these variables.
For a variable $X$, the set of vertices that have directed edges
in $X$ is called the parents of $X$, denoted by $Par(X)$, and the
set of vertices that are endpoints of a directed paths from $X$ is
called the descendants of $X$, denoted by $Des(X)$. A set $V$ of
random variables (on a classical probability space) is said to satisfy
the }}\textit{Causal Markov Condition}\textit{\emph{ relative to a
causal graph }}\textit{$G$}\textit{\emph{ if for any $X\in V$ and
$Y\notin Des(X)$: 
\[
p(X|Par(X),Y)=p(X|Par(X))
\]
}}That is, conditioning on its parents any random variable will be
probabilistically independent from any of its non-descendants. 

Now,\emph{ }\textit{\emph{causal discovery algorithms take as input
a set of conditional and unconditional independence relations among
random variables and provide a causal graph $G$ as output which returns
these independence relations if the Causal Markov Condition is applied
to the graph.}}\footnote{\textit{\emph{More precisely, the independence relations are returned
if all those graphical criteria are applied to the graph which can
be derived from the Causal Markov Condition plus the semi-graphoid
axioms. These criteria are captured by the so-called $d$-separation
criterion (see Pearl, 2009, Ch. 1). }}}\textit{\emph{ Here we do not enter into the details of these algorithms;
rather we simply list the independence relations of the ontological
models of the non-trivial and trivial operational theory and the corresponding
causal graphs.}}\footnote{\textit{\emph{For the application of the causal discovery algorithm
for the EPR-Bell scenario, see }}(Suarez, 2007; Suarez and SanPedro, 2009; Wood
and Spekkens, 2015). \textit{\emph{Also note that the independence
relations also include the ontic states. Thus, the causal discovery
algorithms are not }}\textit{discovery}\textit{\emph{ algorithms in
the sense that they are based solely on the empirically accessible
probabilities. }}}

\vspace{.1in} Let us start with the causal structure of the ontological models
of the \emph{non-trivial} operational theories, $\mathcal{M}$, introduced
in Section \ref{Sec:nontrivi}. The conditional \textit{\emph{independence
relations}} in the ontological models of our three non-trivial theories
are the following: 

\textit{\emph{
\begin{eqnarray}
p(X|A,B) & = & p(X|A)\label{eq:CI1}\\
p(Y|A,B) & = & p(Y|B)\label{eq:CI2}\\
p(X|A,Y,\Lambda) & = & p(X|A,\Lambda)\label{eq:CI3}\\
p(Y|X,B,\Lambda) & = & p(Y|B,\Lambda)\label{eq:CI4}\\
p(X|A,B,\Lambda) & = & p(X|A,\Lambda)\label{eq:CI5}\\
p(Y|A,B,\Lambda) & \stackrel{\tiny\mbox{(CL)}}{=} & p(Y|B,\Lambda)\label{eq:CI6}
\end{eqnarray}
}}

The first two relations are just the non-disturbance equations (\ref{nondist1})-(\ref{nondist2}),
the subsequent relations follow from the appropriate response functions
(\ref{CL-ont21})-(\ref{CL-ont23}), (\ref{EPR-ont21})-(\ref{EPR-ont23}),
and (\ref{PR-ont21})-(\ref{PR-ont23}) of the models specified in
the Appendix. The first five conditional \textit{\emph{independence
relations }}(\ref{eq:CI1})-(\ref{eq:CI5})\textit{\emph{ hold for
all the three models but th}}e last relation (\ref{eq:CI6}) holds
only for the classical model.

\begin{figure}[h]
\centerline{\resizebox{13cm}{!}{\includegraphics{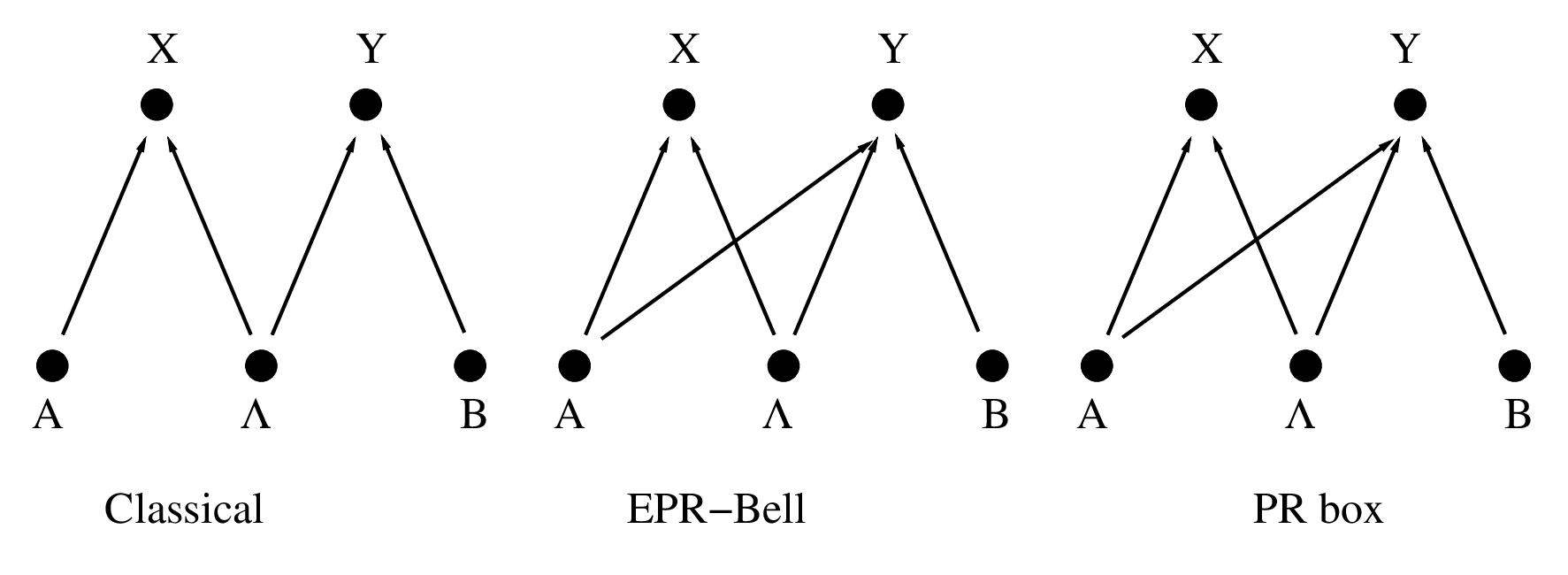}}} \caption{Causal structure of the ontological models with non-trivial compatibility
structure}
\label{Fig:Bridge2}
\end{figure}

The causal graphs which return the independences for the three models
are depicted in\textit{\emph{ }}Figure\ \ref{Fig:Bridge2}.
These graphs are \emph{minimal} in the sense that no subgraph can
return all the \textit{\emph{independence relations}}. Applying the
Causal Markov Condition to the graphs, one obtains also an extra unconditional
\textit{\emph{independence relation}} among the exogenous variables
(that is variables which have no parents):

\begin{equation}
p(A,B,\Lambda)=p(A)p(B)p(\Lambda)\label{eq:nocons}
\end{equation}
These relations are not specified in the model but are consistent
with it. They are a special case of the \emph{no-conspiracy} condition
(\ref{nocons}). 

Observe that there is an edge in the graph of the non-classical models
connecting $A$ and $Y$. This edge represents the causal influence
responsible for simultaneous contextuality: the value of $Y$ causally
depends not only on the value of $X$ and $\Lambda$ but also on the
value of $A$. If $A$ and $Y$ are spacelike separated, this edge
represents a non-local causal influence. Note again, however, that
in constructing the graphs, we relied only on the probabilistic features
of the models and not on the spatiotemporal localizations of the events---in
strong contrast to the usual EPR-Bell analysis.

A further difference between the classical and non-classical models
concerns fine-tuning. To see this, first recall that any joint probability
distribution of the random variables which is \textit{\emph{compatible
with the corresponding causal graphs }}in Figure\ \ref{Fig:Bridge2}
is of the form

\begin{equation}
p(X,Y,A,B,\Lambda)\stackrel{\tiny\mbox{(CL)}}{=}p(X|A,\Lambda)p(Y|B,\Lambda)p(A)p(B)p(\Lambda)\label{joint1}
\end{equation}
for the classical model and of the form

\begin{equation}
p(X,Y,A,B,\Lambda)\stackrel{\tiny\mbox{(EPR, PR)}}{=}p(X|A,\Lambda)p(Y|A,B,\Lambda)p(A)p(B)p(\Lambda)\label{joint2}
\end{equation}
for the non-classical models. In both equations, the conditional probabilities
(the response functions) are called \emph{causal parameters} and the
unconditional probabilities are called \emph{statistical parameters}
(where $p(\Lambda)$ is just a short hand for $p(\Lambda|P)$). By
manipulating these parameters, one obtains all the joint distributions
compatible with the causal graphs. Since causal discovery algorithms
are sensitive only to the independence relations and not to the full
joint probability distribution, the question arises, whether these
independence relations are robust enough against the perturbation
of the causal-statistical parameters, that is whether they continue
to hold when these parameters are not those specified in the Appendix
but take on arbitrary values. If so, the graph is said to be \emph{faithful},
if not, it is said to be \emph{fine-tuned}. 

Now, for the classical model all the conditional independences (\ref{eq:CI1})-(\ref{eq:CI6})
can be derived from the joint probability distribution equation (\ref{joint1})
plus the theorem of total probability. This means that the conditional
independences hold for any choice of the parameters. Thus, the classical
model is faithful. The crucial step in the derivation of the conditional
independences is\textit{\emph{ factorization}}

\[
p(X,Y|A,B,\Lambda)\stackrel{\tiny\mbox{(CL)}}{=}p(X|A,\Lambda)p(Y|B,\Lambda)
\]
By summing up for the different variables, one recovers the different
conditional independences (\ref{eq:CI1})-(\ref{eq:CI6}). In the
non-classical models, however, one has

\[
p(X,Y|A,B,\Lambda)\stackrel{\tiny\mbox{(EPR, PR)}}{=}p(X|A,\Lambda)p(Y|A,B,\Lambda)
\]
instead of the factorization and hence summing up does \emph{not}
recover (\ref{eq:CI6}), (\ref{eq:CI4}) and the non-disturbance (\ref{eq:CI2}).
And indeed, for a non-zero measure of the parameters, these conditional
independences will fail to hold. Therefore, the non-classical models
are fine-tuned. 

These facts are in tune with Cavalcanti's (2018) theorem on bipartite
Bell scenarios stating that every causal model for a non-disturbing
operational theory violating the CHSH inequality requires fine-tuning.
Cavalcanti's result highlights a deep connection between simultaneous
contextuality of the model and fine tuning of the corresponding graph.
At the end of his paper, he asks whether also \emph{measurement} noncontextuality
can be understood as arising from the no-fine-tuning condition. 

\begin{figure}[h]
\centerline{\resizebox{2.5cm}{!}{\includegraphics{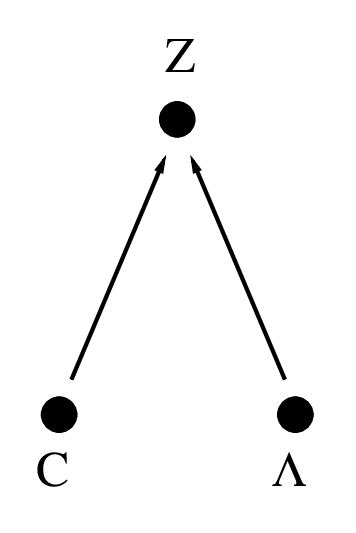}}} \caption{Causal structure of the ontological models with trivial compatibility
structure}
\label{Fig:Bridge2-1}
\end{figure}

To answer Cavalcanti's question, \textit{\emph{let us now turn to
the causal structure of the ontological models of the }}\textit{trivial
theory}\textit{\emph{. In these models there are no conditional independence
relations, except}} among the exogenous variables:

\begin{equation}
p(C,\Lambda)=p(C)p(\Lambda)\label{eq:nocons2}
\end{equation}
which is again consistent with the models. The causal graph which
is compatible with (\ref{eq:nocons2}) is depicted in\textit{\emph{
}}Figure\ \ref{Fig:Bridge2-1}.
Note, that the graph is the same for all three models. The four measurements
cannot be simultaneously performed, therefore the models are (trivially)
simultaneously noncontextual. The models are also faithful since any
choice of the parameters in the joint probability distribution equation 

\begin{equation}
p(Z,C,\Lambda)=p(Z|C,\Lambda)p(C)p(\Lambda)\label{joint3}
\end{equation}
compatible with the graph in Figure\ \ref{Fig:Bridge2-1} will return
the same independence relations, that is (\ref{eq:nocons2}). Thus,
if Cavalcanti's question is whether measurement contextual ontological
models for an operational theory are also fine-tuned, then the answer
is no. Both the models of the non-trivial and trivial non-classical
operational theories are measurement contextual, still the causal
graphs are fine-tuned for the former and faithful for the latter.
Fine-tuning relates to simultaneous contextuality but not to measurement
contextuality. 

To sum up, the causal graph of the model of the classical and non-classical
non-trivial operational theories are different; the graphs of the
non-classical models are fine-tuned and contain a directed edge representing
the causal connection responsible for simultaneous contextuality.
This difference between the graphs collapses upon trivializing the
theories; the graph of all three models will be the same: trivial
and faithful. 

\section{How trivialization leads to trivial causal graphs?}

\label{Sec:Reptrivi}

In this section, I return to the concept of trivialization and show
how it leads to trivial causal graphs. At the beginning of Section
\ref{Sec:trivi}, we replaced the non-trivial operational theory 

\[
\mathcal{M_{\mbox{}}=}\{A_{0},A_{1},B_{0},B_{1},\,A_{0}\wedge B_{0},\,A_{0}\wedge B_{1},\,A_{1}\wedge B_{0},\,A_{1}\wedge B_{1}\}
\]
with an operationally equivalent but \emph{different} theory:

\[
\mathcal{M}'=\{C_{00},C_{01},C_{10},C_{11}\}
\]

Why are the causal graphs different for $\mathcal{M}$ and $\mathcal{M}'$?

As stated in the previous section, causal graphs arise from the conditional
independences of an ontological model via the causal discovery algorithms.
Let us pick one such independence from the non-trivial theory, for
example, (\ref{eq:CI5}) with $A=0$:

\begin{equation}
p(X|A_{0}\wedge B_{0},\Lambda)=p(X|A_{0}\wedge B_{1},\Lambda)=p(X|A_{0},\Lambda)\label{eq:CI5a}
\end{equation}

What would be the ``analogue'' of (\ref{eq:CI5a}) in the trivial
theory? 

One has two options---unfortunately neither yielding a conditional
independence. The first option is to replace $A_{0}\wedge B_{0}$
by $C_{00}$, $A_{0}\wedge B_{1}$ by $C_{01}$ and $A_{0}$ by $C_{00}^{(1)}$
or $C_{01}^{(1)}$. Thus, we obtain:

\begin{equation}
p(Z_{1}|C_{00},\Lambda)=p(Z_{1}|C_{01},\Lambda)=p(Z_{1}|C_{00}^{(1)},\Lambda)=p(Z_{1}|C_{01}^{(1)},\Lambda)\label{eq:CI5b}
\end{equation}
But (\ref{eq:CI5b}) contains no conjunction, only marginalization.
Consequently, (\ref{eq:CI5b}) is not a conditional independence.
The other option is to replace, say, $A_{0}$ by $C_{00}^{(1)}$,
$B_{0}$ by $C_{10}^{(2)}$ and $B_{1}$ by $C_{11}^{(2)}$. We obtain
\begin{equation}
p(Z_{1}|C_{00}^{(1)}\wedge C_{10}^{(2)},\Lambda)=p(Z_{1}|C_{00}^{(1)}\wedge C_{11}^{(2)},\Lambda)=p(Z_{1}|C_{00}^{(1)},\Lambda)\label{eq:CI5c}
\end{equation}
The problem with (\ref{eq:CI5c}) is that it is ill-defined: the conjunctions
$C_{00}^{(1)}\wedge C_{10}^{(2)}$ and $C_{00}^{(1)}\wedge C_{11}^{(2)}$
are not defined since $C_{00}$, $C_{10}$ and $C_{11}$ cannot be
simultaneously measured. And this is as it should be since $\mathcal{M}'$
is a trivial theory.

In sum, trivialization removes conditional independences because in
the trivialized theory there will be no simultaneous measurements.
Consequently, trivialization leads to trivial causal graphs.

An anonymous referee suggested the following recovery of the conditional
independence (\ref{eq:CI5a}). First, allow for an operational theory
to contain measurements which are defined not only by \emph{conjunctions}
but also by \emph{disjunctions} of instructions. That is, allow for
measurements like, for example, $C_{00}\vee C_{10}$. Now, consider
the following operational theory
\[
\mathcal{M}''=\{C_{00}\vee C_{01},C_{11}\vee C_{01},C_{11}\vee C_{10},C_{00}\vee C_{10},\,C_{00},C_{01},C_{10},C_{11}\}
\]
Since $C_{00}$ is a conjunction of $C_{00}\vee C_{01}$ and $C_{00}\vee C_{10}$
(and similarly for $C_{01},C_{10},C_{11}$), $\mathcal{M}''$ will
be a nontrivial theory. Then the conditional independence (\ref{eq:CI5a})
will look like this: 

\begin{equation}
p(Z_{1}|(C_{00}\vee C_{01})\wedge(C_{00}\vee C_{10}),\Lambda)=p(Z_{1}|(C_{00}\vee C_{01})\wedge(C_{11}\vee C_{01}),\Lambda)=p(Z_{1}|C_{00}\vee C_{01},\Lambda)\label{eq:CI5c-1}
\end{equation}
Equation (\ref{eq:CI5c-1}) indeed recovers (\ref{eq:CI5a}); so embracing
disjunctions seems to bring back non-trivial theories and non-trivial
causal structure. However, the operational theory $\mathcal{M}''$
is different from $\mathcal{M}'$. Namely, an operational theory is
defined by a set of measurements (and preparations). If one extends
the set of measurements either by conjunctions or by disjunctions,
one gets \emph{another} operation theory. This extended theory can
well have a causal structure which is different from that of the original
theory. But even if $\mathcal{M}''$ has a non-trivial causal structure,
this does not invalidate our central claim that the causal structure
of $\mathcal{M}'$ is trivial while the causal structure of $\mathcal{M}$
which is operationally equivalent to $\mathcal{M}'$ is non-trivial. 

To be clear, I have no problem with allowing for \textit{disjunctive} measurements (``measure polarization along axis $x$ \emph{or} measure polarization along axis $y$``) in an operational theory, as in the above example. What is important, however, is to keep in mind that in the definition of simultaneous contextuality (\ref{NC1}) \textit{only conjunctions} pop up. Therefore, the dividing line draws between theories in which there are only basic measurements---be them disjunctive or not---and theories in which there are also conjunctions thereof.  The causal structure of the former theories will always be trivial, whereas that of the latter can be nontrivial. And this difference depends only on the presence of conjunctions and not of disjunctions in the operational theory.

It is worths reflecting for a moment on the difference between trivial
and non-trivial theories from a general Bridgmanian perspective. If
we trivialize a theory, we change the empirical content. The new basic
measurements will not be the same as the old maximally simultaneous
measurements and the old basic measurements will not be the same of
the marginalization of the new basic measurements. As a special consequence
of this general fact, measurements sitting in two different contexts
in the non-trivial theory (as $A_{0}$ in our above example) will
``multiply realized'' in the trivial theory by two different marginalized
measurements (by $C_{00}^{(1)}$ and $C_{01}^{(1)}$ in our example).
This results in the disappearing of those conditional independences
in which the original measurement was featuring and consequently in
the a radical change of the causal structure based on these conditional
independences. From a general point of view, this is understandable:
operational theories with different empirical content can have different
causal explanation. A causal explanation relies not only on the outcome
statistics of the measurements but also on their compatibility structure.
Focusing only on operationally equivalent measurement classes, this
information about simultaneous measurability gets lost.

\section{Quantum mechanics}

\label{Sec:QMech}

Quantum mechanics, at least in the minimalist interpretation, is an
operational theory in a special linear algebraic representation. Therefore,
it is instructive to see how quantum mechanics represents the EPR-Bell
scenario and how this representation relates to the Bridgmanian and
the standard identification of observables. The probabilities of the
both the non-trivial operational theory (\ref{OP}) and (\ref{EPR})
and the trivial operational theory (\ref{EPR-1}) are generated quantum
mechanically as follows: 

\begin{eqnarray}
\braket{\Psi_{s}\vert(\mathbf{X^{A}\otimes{\normalcolor \mathbf{I}}})\Psi_{s}} & = & p(X|A,P_{\tiny\mbox{EPR}})\label{eq:QM1}\\
\braket{\Psi_{s}\vert\mathbf{(I\otimes Y}^{\mathbf{B}})\Psi_{s}} & = & p(Y|B,P_{\tiny\mbox{EPR}})\label{eq:QM2}\\
\braket{\Psi_{s}\vert(\mathbf{X^{A}\otimes Y^{B}})\Psi_{s}} & = & p(X,Y|A,B,P_{\tiny\mbox{EPR}})\label{eq:QM3}
\end{eqnarray}
where $\ket{\Psi_{s}}$ is the singlet state representing the preparation
$P_{\tiny\mbox{EPR}}$ in the Hilbert space $H_{2}\otimes H_{2}$;
$\mathbf{I}$ is the unit operator in $H_{2}$; and $\mathbf{X}^{\mathbf{A}}$
and $\mathbf{Y}^{\mathbf{B}}$ scroll over eight projections 
\begin{eqnarray*}
\mathbf{X_{0}}^{\mathbf{A_{0}}},\,\mathbf{X_{1}}^{\mathbf{A_{0}}},\,\mathbf{X_{0}}^{\mathbf{A_{1}}},\,\mathbf{X_{1}}^{\mathbf{A_{1}}}\\
\mathbf{Y_{0}}^{\mathbf{B_{0}}},\,\mathbf{Y_{1}}^{\mathbf{B_{0}}},\,\mathbf{Y_{0}}^{\mathbf{B_{1}}},\,\mathbf{Y_{1}}^{\mathbf{B_{1}}}
\end{eqnarray*}
 corresponding to eight unit vectors $\ket{X^{A}}$ and $\ket{Y^{B}}$
in $H_{2}$ such that 
\begin{eqnarray*}
|\!\braket{X^{A}\vert Y^{B}}\!|^{2} & = & \begin{cases}
\begin{array}{cc}
\frac{3}{4} & \quad\mbox{if}\,\,X\oplus Y=0\,\,\mbox{and}\,\,A\cdot B=0\\
\frac{1}{4} & \quad\mbox{if}\,\,X\oplus Y=1\,\,\mbox{and}\,\,A\cdot B=0\\
1 & \quad\mbox{if}\,\,X\oplus Y=0\,\,\mbox{and}\,\,A\cdot B=1\\
0 & \quad\mbox{if}\,\,X\oplus Y=1\,\,\mbox{and}\,\,A\cdot B=1
\end{array}\end{cases}
\end{eqnarray*}
The operators representing the four measurements are:

\begin{eqnarray*}
\mathbf{A_{0}=X_{0}}^{\mathbf{A_{0}}}-\mathbf{X_{1}}^{\mathbf{A_{0}}},\quad &  & \mathbf{A_{1}=X_{0}}^{\mathbf{A_{1}}}-\mathbf{X_{1}}^{\mathbf{A_{1}}}\\
\mathbf{B_{0}=Y_{0}}^{\mathbf{B_{0}}}-\mathbf{Y_{1}}^{\mathbf{B_{0}}},\quad &  & \mathbf{B_{1}=Y_{0}}^{\mathbf{B_{1}}}-\mathbf{Y_{1}}^{\mathbf{B_{1}}}
\end{eqnarray*}
with eigenvalues $\pm1$. 

The operators, however, represent different measurements in the non-trivial
and trivial operational theory. Consider, for example, the quantum
optical realization of the EPR-Bell scenario. In both operational
theories, one prepares an ensemble of photon pairs in singlet state
and performs certain polarization measurements on the pairs. 

In the \emph{non-trivial} theory, one has four \emph{local} measurements:
two linear polarization measurements on the left photon, $A_{0}$
and $A_{1}$, and two linear polarization measurements on the right
photon, $B_{0}$ and $B_{1}$. These measurements are the following:
\begin{description}
\item [{$A_{0}:$}] Measure the linear polarization of the left photon
along a given transverse axis $a_{0}$ (with outcome $+1$ if the
photon passes the polarizer and $-1$ if not)
\item [{$A_{1}:$}] Measure the linear polarization of the left photon
along a transverse axis $a_{1}$ at $60^{\circ}$ from the axis $a_{0}$
\item [{$B_{0}:$}] Measure the linear polarization of the right photon
along a transverse axis $b_{0}$ at $60^{\circ}$ from the axis both
$a_{0}$ and $a_{1}$
\item [{$B_{1}:$}] Measure the linear polarization of the right photon
along the transverse axis $b_{1}=a_{1}$
\end{description}
The polarization measurements on the left subsystem can be simultaneously
performed with the polarization measurements on the right subsystem
realizing the simultaneous measurements $A_{0}\wedge B_{0}$, $A_{0}\wedge B_{1}$,
$A_{1}\wedge B_{0}$, and $A_{1}\wedge B_{1}$. The local measurements
do not disturb one another, still the ontological model constructed
above is simultaneous contextual: performing measurement $A_{0}$
or $A_{1}$ causally influences the outcomes of $B_{0}$ and $B_{1}$.
Since the events $A$ and $Y$ are spacelike separated, this is a
clear violation of local causality. 

In the \emph{trivial} operational theory, we replace the local measurements
with \emph{global} measurements. We will have four new measurements,
each with four outcomes represented by four orthogonal unit vectors
in $H_{2}\otimes H_{2}$:
\begin{description}
\item [{$C_{00}:$}] Perform a global polarization measurement on the photon
pair with four outcomes corresponding to the basis $\left\{ \ket{X_{0}^{A_{0}}}\otimes\ket{Y_{0}^{B_{0}}},\ket{X_{0}^{A_{0}}}\otimes\ket{Y_{1}^{B_{0}}},\ket{X_{1}^{A_{0}}}\otimes\ket{Y_{0}^{B_{0}}},\ket{X_{1}^{A_{0}}}\otimes\ket{Y_{1}^{B_{0}}}\right\} $
\item [{$C_{01}:$}] Perform a global polarization measurement on the photon
pair with four outcomes corresponding to the basis $\left\{ \ket{X_{0}^{A_{0}}}\otimes\ket{Y_{0}^{B_{1}}},\ket{X_{0}^{A_{0}}}\otimes\ket{Y_{1}^{B_{1}}},\ket{X_{1}^{A_{0}}}\otimes\ket{Y_{0}^{B_{1}}},\ket{X_{1}^{A_{0}}}\otimes\ket{Y_{1}^{B_{1}}}\right\} $
\item [{$C_{10}:$}] Perform a global polarization measurement on the photon
pair with four outcomes corresponding to the basis $\left\{ \ket{X_{0}^{A_{1}}}\otimes\ket{Y_{0}^{B_{0}}},\ket{X_{0}^{A_{1}}}\otimes\ket{Y_{1}^{B_{0}}},\ket{X_{1}^{A_{1}}}\otimes\ket{Y_{0}^{B_{0}}},\ket{X_{1}^{A_{1}}}\otimes\ket{Y_{1}^{B_{0}}}\right\} $
\item [{$C_{11}:$}] Perform a global polarization measurement on the photon
pair with four outcomes corresponding to the basis $\left\{ \ket{X_{0}^{A_{1}}}\otimes\ket{Y_{0}^{B_{1}}},\ket{X_{0}^{A_{1}}}\otimes\ket{Y_{1}^{B_{1}}},\ket{X_{1}^{A_{1}}}\otimes\ket{Y_{0}^{B_{1}}},\ket{X_{1}^{A_{1}}}\otimes\ket{Y_{1}^{B_{1}}}\right\} $
\end{description}
Note that these global polarization measurements are realized by a
complicated arrangement of beam splitters, polarizing beam splitters,
wave plates, photo detectors and other non-linear optical devices
(Mattle et al., 1996; L\"utkenhaus et al., 1999; Weihs and Zeilinger
2001). What is important, is that $C_{00}$ is \emph{not} simply performing
a linear polarization measurement on the left photon along axis $a_{0}$
and performing a linear polarization measurement on the right photon
along a given transverse axis $b_{0}$. In other words, $C_{00}$
is \emph{not} the same measurement as $A_{0}\wedge B_{0}$; they are
only operationally equivalent. Consequently, $C_{00}^{(1)}$ will
not be the same as $A_{0}$; they will be only operationally equivalent.

This new operational theory has a trivial compatibility structure:
$C_{01}$ and $C_{11}$ cannot be performed simultaneously, that is,
they cannot be performed on the same pair of photons. Consequently,
any ontological model for the theory is (trivially) simultaneously
noncontextual. But the model we provided will be measurement contextual:
some ontic states will provide different outcomes for the $C_{01}^{(2)}$
and $C_{11}^{(2)}$ contrary to the fact that they are operationally
equivalent. Note, however, that measurement contextuality does not
lead to the violation of local causality. 

In the Introduction, we discerned the Bridgmanian and the standard
identification of observables. In the first case, we identified observables
with operators, in the second, with measurements. Applying this distinction
to the EPR-Bell scenario, one gets the following schema: 
\begin{center}
\begin{tabular}{rccccccc} &&&&&&&\\ 
  &  & \textit{Bridgmanian} &  &  &  & \textit{Standard} &  \\
  &   &  &  &  &  &  &  \\ 
\textit{Operator:} \quad &  & $\bf{A_0}$ &  &  &  & $\bf{A_0}$ &  \\
  &   &  $\swarrow \quad \searrow$ &   &  &  & $\downarrow$ &  \\  
\textit{Observable:} \quad & & $\mathcal{O}_1 \quad \quad \quad \mathcal{O}_2$   &  &  &  & $\mathcal{O}$ &  \\
  &  & $\downarrow \quad \quad \quad \quad \downarrow$   &  &  &  & $\swarrow \quad \searrow$ & \\  
\textit{Measurement:} \quad & & $ A_0 \quad \sim \quad C^{(2)}_{00}$ &  &  & & $\quad A_0 \quad \sim \quad C^{(2)}_{00}$ & \\
  &  &  & &  &  &  & 
\end{tabular}
\par\end{center}

The local and global measurements are represented by the same operator
in quantum mechanics. But do they measure the same observable? According
to the standard approach: yes; according to the Brigdmannian approach:
no.

\section{Conclusions}

\label{Sec:Conclu}

Operational quantum mechanics is a special operational theory in a
linear algebraic representation. A distinctive feature of this theory
is operational equivalence, the representation of different (and not
necessarily simultaneously performable) measurements providing the
same outcome statistics in every quantum state by the same self-adjoint
operator (or POVM). From the perspective of strict operationalism,
the identity of the representation of such measurements does not mean
the identity of the measured observables. In this paper, I intended
to explore some of the consequences of this Bridgmanian perspective
in quantum theory and in general operational theories. We saw, how
certain essential properties of the underlying ontological models
changed if some measurements were replaced by other operationally
equivalent measurements. This change is a straightforward consequence
of the sensitivity of the ontological models to the compatibility
structure of the theory. To illustrate this change in quantum mechanics,
I took the example of the EPR-Bell scenario and compared the ontological
models of the non-trivial and the trivial operational theories realizing
the EPR-Bell scenario by local and global measurements, respectively.
The EPR-Bell situation, however, was not peculiar whatsoever; we could
have equally well used the GHZ or the Peres-Mermin case to this goal.
The four commuting operators in the horizontal line of the GHZ pentagram
\[
\boldsymbol{\sigma}_{z}\otimes\boldsymbol{\sigma}_{z}\otimes\boldsymbol{\sigma}_{z}\quad\quad\boldsymbol{\sigma}_{z}\otimes\boldsymbol{\sigma}_{x}\otimes\boldsymbol{\sigma}_{x}\quad\quad\boldsymbol{\sigma}_{x}\otimes\boldsymbol{\sigma}_{z}\otimes\boldsymbol{\sigma}_{x}\quad\quad\boldsymbol{\sigma}_{x}\otimes\boldsymbol{\sigma}_{x}\otimes\boldsymbol{\sigma}_{z}
\]
or the three commuting operators in the third column of the Peres-Mermin
square 
\[
\boldsymbol{\sigma}_{z}\otimes\boldsymbol{\sigma}_{z}\quad\quad\boldsymbol{\sigma}_{y}\otimes\boldsymbol{\sigma}_{y}\quad\quad\boldsymbol{\sigma}_{x}\otimes\boldsymbol{\sigma}_{x}
\]
can also be represented both by local measurements on individual photons
(represented by the graphs in the Figure\ \ref{Fig:Bridge_graph1})
and also by complicated global GHZ or Bell state measurements on photon
pairs or triples (represented by the linegraphs in the Figure\ \ref{Fig:Bridge_linegraph1}).
These local and global measurements are different and so are the ontological
models. All ontological models will be measurement contextual, but
those for global measurements will be simultaneously noncontextual
and will have a trivial causal structure (Hofer-Szab\'o, 2021a, b, 2022).
All these results point in the same direction which is also the main
message of this paper: Operationally equivalent families of measurements
represented by the same operators in quantum mechanics can give rise
to ontological models with highly different features. Thus, to study
these models, it is not enough to simply investigate quantum mechanics
at an abstract mathematical level; we also need to take into consideration
the measurements represented by the operators. This is the lesson
that we can learn from Bridgman. 

\section*{Appendix}

\textit{\emph{Three outcome-deterministic ontological model for the
three operational theories with non-trivial compatibility structure:}}

\vspace{.1in} \emph{Classical theory. }
\begin{itemize}
\item Set of ontic states: \textit{\emph{$\mathcal{L}=\{\Lambda_{0},\Lambda_{1}\}$}} 
\item Random variable on \textit{\emph{$\mathcal{L}$ }}: $\Lambda=0,1$
\item Probability distribution: 
\end{itemize}
\begin{eqnarray}
p(\Lambda|P_{\tiny\mbox{CL}}) & = & \frac{1}{2}\label{CL-ont1}
\end{eqnarray}

\begin{itemize}
\item Response functions of the non-trivial theory: 
\begin{eqnarray}
p(X|A,\Lambda) & = & \delta_{X,\Lambda}\label{CL-ont21}\\
p(Y|B,\Lambda) & = & \delta_{Y,\Lambda}\label{CL-ont22}\\
p(X,Y|A,B,\Lambda) & = & \delta_{X,\Lambda}\cdot\delta_{Y,\Lambda}\label{CL-ont23}
\end{eqnarray}
where $\delta$ is the Kronecker delta function.
\item Response functions of the trivial theory: 
\begin{eqnarray}
p(Z|C,\Lambda) & = & \delta_{Z_{1},\Lambda}\cdot\delta_{Z_{2},\Lambda}\label{CL-ont24}
\end{eqnarray}
\end{itemize}
\vspace{.1in}\emph{ The EPR-Bell scenario.}
\begin{itemize}
\item Set of ontic states: \textit{\emph{$\mathcal{L}\times\mathcal{L}$}}
where $\mathcal{L}=\{\Lambda_{0},\Lambda_{1}\}$
\item Random variable on \textit{\emph{$\mathcal{L}\times\mathcal{L}$}}
: $\Lambda_{1}\times\Lambda_{2}$ with $\Lambda_{1},\Lambda_{2}=0,1$ 
\item Probability distribution: 
\end{itemize}
\begin{eqnarray}
p(\Lambda_{1},\Lambda_{2}|P_{\tiny\mbox{EPR}}) & = & \begin{cases}
\begin{array}{cc}
\frac{1}{8} & \quad\mbox{if}\,\,\Lambda_{1}\oplus\Lambda_{2}=1\\
\frac{3}{8} & \mbox{otherwise}
\end{array}\end{cases}\label{EPR-ont1-1}
\end{eqnarray}

\begin{itemize}
\item Response functions of the non-trivial theory: 
\begin{eqnarray}
p(X|A,\Lambda_{1},\Lambda_{2}) & = & \delta_{X,\Lambda_{1}}\label{EPR-ont21}\\
p(Y|B,\Lambda_{1},\Lambda_{2}) & = & \delta_{Y,\Lambda_{2}}\label{EPR-ont22}\\
p(X,Y|A,B,\Lambda_{1},\Lambda_{2}) & = & \delta_{X,\Lambda_{1}}\cdot(\delta_{Y\oplus(A\cdot B),\Lambda_{2}}\cdot\delta_{\Lambda_{1}\oplus\Lambda_{2},1}+\delta_{Y,\Lambda_{2}}\cdot\delta_{\Lambda_{1}\oplus\Lambda_{2},0})\label{EPR-ont23}
\end{eqnarray}
\item Response functions of the trivial theory: 
\begin{eqnarray}
p(Z|C,\Lambda) & = & \delta_{Z_{1},\Lambda_{1}}\cdot(\delta_{Z_{2}\oplus(C_{1}\cdot C_{2}),\Lambda_{2}}\cdot\delta_{\Lambda_{1}\oplus\Lambda_{2},1}+\delta_{Z_{2},\Lambda_{2}}\cdot\delta_{\Lambda_{1}\oplus\Lambda_{2},0})\label{EPR-ont24}
\end{eqnarray}
\end{itemize}
\vspace{.1in} \emph{PR box.}
\begin{itemize}
\item Set of ontic states: \textit{\emph{$\mathcal{L}=\{\Lambda_{0},\Lambda_{1}\}$}} 
\item Random variable on \textit{\emph{$\mathcal{L}$ }}: $\Lambda=0,1$
\item Probability distribution: 
\end{itemize}
\begin{eqnarray}
p(\Lambda|P_{\tiny\mbox{PR}}) & = & \frac{1}{2}\label{PR-ont1-1}
\end{eqnarray}

\begin{itemize}
\item Response functions of the non-trivial theory: 
\end{itemize}
\begin{eqnarray}
p(X|A,\Lambda) & = & \delta_{X,\Lambda}\label{PR-ont21}\\
p(Y|B,\Lambda) & = & \delta_{Y,\Lambda}\label{PR-ont22}\\
p(X,Y|A,B,\Lambda) & = & \delta_{X,\Lambda}\cdot\delta_{Y\oplus(A\cdot B),\Lambda}\label{PR-ont23}
\end{eqnarray}

\begin{itemize}
\item Response functions of the trivial theory:
\end{itemize}
\begin{eqnarray}
p(Z|C,\Lambda) & = & \delta_{Z_{1},\Lambda}\cdot\delta_{Z_{2}\oplus(C_{1}\cdot C_{2}),\Lambda}\label{eq:PR-ont24}
\end{eqnarray}

\textbf{Acknowledgements.} This work has been supported by the Friedrich
Wilhelm Bessel Research Award of the Alexander von Humboldt Foundation,
the Hungarian National Research, Development and Innovation Office
(K-134275). I would like to thank for the two anonymous reviewers
for their valuable comments.

\section*{References}

\footnotesize

\begin{list} { }{\setlength{\itemindent}{-15pt} \setlength{\leftmargin}{15pt}}\par 

\item Abramsky S., and A. Brandenburger, (2011). The sheaf-theoretic
structure of non-locality and contextuality, New J. Phys, 13, 113036.\par 

\item Abramsky, S., R. S. Barbosa, K. Kishida, R. Lal, S. Mansfield,
(2017). Contextuality, cohomology and paradox, URL = https://arxiv.org/abs/1502.03097.\par 

\item Bridgman, P. W. (1958). The Logic of Modern Physics (New York:
The Macmillan Company).

\item  Busch P.,  Lahti  P., Pellonp\"a\"a J-P., Ylinen K. (2016). Quantum Measurement. Springer. \par 

\item Cavalcanti, E. (2018). Classical Causal Models for Bell and
Kochen-Specker Inequality Violations Require Fine-Tuning, Phys. Rev.
X, 8, 021018.\par 

\item Chang, H. (2019). Operationalism, Stanford Encyclopedia of
Philosophy. URL = https://plato.stanford.edu/entries/ operationalism.
\par 

\item Clauser, J. F., M.A. Horne, A. Shimony and R. A. Holt, (1969).
Proposed experiment to test local hidden-variable theories, Phys.
Rev. Lett., 23, 880-884.\par 

\item Glymour, C., Scheines, R., and Spirtes, P. (2001). Causation,
Prediction, and Search, (Cambridge: The MIT Press).\par 

\item Greenberger, D. M., Horne, M. A., Shimony, A. and Zeilinger,
A. (1990). Bell?s theorem without inequalities, Am. J. Phys. 58, 1131
\par 

\item  D'Ariano G. M.,  Chiribella G., Perinotti P. (2017). Quantum Theory from First Principles: An Informational Approach, Cambridge: Cambridge University Press. \par 

\item Held, C. (2022). The Kochen-Specker Theorem, Stanford Encyclopedia
of Philosophy \par 

%\item Hofer-Szab\'o, G. (2021a). Commutativity, comeasurability, and contextuality in the Kochen-Specker arguments, Phil. Sci., 88, 483-510.\par 

%\item Hofer-Szab\'o, G. (2021b). Three noncontextual hidden variable models for the Peres-Mermin square, Eur. J. Phil. Sci., 11, 30.\par 

%\item Hofer-Szab\'o, G. (2022). Two concepts of noncontextuality in quantum mechanics, Studies in History and Philosophy of Science, 93, 21-29. .\par 

\item Kochen, S., and E. P. Specker (1967). The problem of hidden
variables in quantum mechanics, J. Math. Mech., 17, 59--87.\par 

\item Lin, Y. (2021). Conspiracy in ontological models: $\lambda$
sufficiency and measurement contextuality, Phys. Rev. A 103, 022211.\par 

\item Ludwig, G. (1983). Foundations of Quantum Mechanics. Springer. \par 

\item L\"utkenhaus, N., Calsamiglia, J., and Suominen, K-A. (1999).
On Bell measurements for teleportation, Phys. Rev. A, 59, 3295.\par 

\item  Mackey G. W. (1957). Quantum Mechanics and Hilbert Space. The American Mathematical Monthly, 64, 45-57. \par 

\item Mattle, K., Weinfurter, H., Kwiat, P. G., and Zeilinger A.
(1996). Dense Coding in Experimental Quantum Communication, Phys.
Rev. Lett, 76 (25), 4656-4659. \par 

\item {\footnotesize{} Mermin, D. (1993). Ontological states and
the two theorems of John Bell, Rev. Mod. Phys., 65 (3), 803-815. }\par 

\item {\footnotesize{} }Pearl, J. (2009). Causality: Models, Reasoning,
and Inference, (Cambridge: Cambridge University Press)\par 

\item {\footnotesize{} Peres, A. (1990). Incompatible Results of
Quantum Measurements,? Phys. Lett. A, 151, 107-108. }\par 

\item Popescu, S., and D. Rohrlich. (1994). Nonlocality as an axiom, Found. Phys., 24, 379-385.\par 

\item Reichenbach, H. (1927). The Philosophy of Space and Time, New York: Dover Publications, 1958.\par 

\item Schlick, M. (1930 {[}1979{]}). On the Foundations of Knowledge,
in Philosophical Papers, vol. 2 (1925--1936), H. L. Mulder and B.
F. B. van de Velde-Schlick (eds.), Dordrecht: Reidel, pp. 370--387.
\par 

\item Spekkens, R. W. (2005). Contextuality for preparations transformations
and unsharp measurements, Phys. Rev. A 71:052108.\par 

\item Suarez, M. (2007). Causal Inference in Quantum Mechanics: A Reassessment, in: F. Russo and J. Williamson (eds.), Causality and Probability in the Sciences. College Publications. \par 

\item Suarez, M. and SanPedro, I. (2009). Causal Markov, Robustness
and the Quantum Correlations, in: M. Suarez (ed.), Probabilities,
Causes and Propensities in Physics, Synthese Library, Springer, Ch.
8., 173-193. \par 

\item Suppes, P. (1951). A set of independent axioms for extensive quantities, Portugaliae Mathematica, 10(4): 163–172. \par 

\item Tal, E. (2020). Measurement in Science, Stanford Encyclopedia of
Philosophy. URL = https://plato.stanford.edu/entries/measurement-science \par 

\item Trout, J.D. (1998). Measuring the intentional world: Realism, naturalism, and quantitative methods in the behavioral sciences, Oxford: Oxford University Press. \par 

\item Weihs, G., and Zeilinger, A. (2001). Photon statistics at beam
splitters: an essential tool in quantum information and teleportation,
in: Jan Perina (ed), Coherence and Statistics of Photons and Atoms,
(John Wiley and Son, Inc, New York) 262-288. \par 

\item van Fraassen, B.C. (1980). The Scientific Image, Oxford: Clarendon Press. \par 

\item von Neumann, J. (1932). Mathematische Grundlagen der Quantenmechanik, Springer. \par 

Mathematische Grundlagen der Quantenmechanik

\item Wood, C. J., and Spekkens, R. W. (2015). The lesson of causal
discovery algorithms for quantum correlations: causal explanations
of Bell-inequality violations require fine-tuning, New J. Phys. 17,
033002.\par \end{list}

\end{document}